\documentclass[conference]{IEEEtran}

\usepackage[linesnumbered,titlenotnumbered,boxed]{algorithm2e}
\usepackage{mathtools}
\usepackage{multirow}
\usepackage{caption}
\usepackage{subcaption}
\usepackage{balance}

\usepackage{cite}
\usepackage{amsmath,amssymb,amsfonts,amsbsy}
\usepackage{algorithmic}
\usepackage{graphicx}
\usepackage{textcomp}
\usepackage{xcolor}

\usepackage{ifsym}
\usepackage{latexsym}
\usepackage{array}
\usepackage[inline]{enumitem}
\usepackage{listings}
\usepackage{relsize}
\usepackage{adjustbox}
\usepackage{colortbl}
\usepackage{ifthen}
\usepackage{cancel}
\usepackage{cleveref}
\usepackage{scalerel}
\usepackage{mdframed}

\usepackage{tikz}
\usetikzlibrary{arrows,calc, patterns, shapes, positioning, matrix}
\tikzset{
	>=stealth',
	help lines/.style={dashed, thick},
	axis/.style={<->},
	important line/.style={thick},
	connection/.style={thick, dotted},
}
\usepackage{pgfplots}
\pgfplotsset{compat=newest}
\usepgfplotslibrary{groupplots}

\definecolor{dkgreen}{rgb}{0,0.6,0}
\definecolor{darkred}{rgb}{0.8,0,0}
\definecolor{darkblue}{rgb}{0,0,0.6}
\definecolor{lightBlue}{RGB}{0,113,202}
\definecolor{lightBlue2}{RGB}{0,167,227}
\definecolor{mauve}{rgb}{0.58,0,0.82}
\definecolor{orangeDark}{RGB}{255,128,0}

\newcommand{\mathcoloring}[2]{\mathbin{\textcolor{#1}{#2}}}

\newcommand{\crc}[1]{#1}

\newenvironment{centermath}
{\begin{center}$\displaystyle}
	{$\end{center}}

\newcommand{\coordSystem}[2]{
	\coordinate (y) at (0,#2);
	\coordinate (x) at (#1,0);
	\draw[->] (0,0) --  (x);
	\draw[->] (0,0) -- (y);
	\draw[help lines, color=black!50] (0,0) -- (#2,#2);
	\draw (#1+0, 0) node[right] {$\tcurr$};
	\draw (-0.1, #2-0) node[above] {$vt$};
}

\newcommand{\strLineColor}[4]{
	\draw[thick,#4] (#1, #2) -- (#3, #2);
}

\newcommand{\strLineColorY}[4]{
	\draw[#4,ultra thick] (#1, #2) -- (#1, #3);
	\draw[#4,ultra thick] (#1 -0.2, #2) -- (#1 + 0.2, #2);
	\draw[#4,ultra thick] (#1 -0.2, #3) -- (#1 + 0.2, #3);

}

\newcommand{\strNowColor}[6]{
	\draw[thick, #6] (#1, #2) -- (#3, #2);
	\draw[thick, #6] (#3, #2) -- (#4, #5);
}

\newcommand{\strUpNowColor}[6]{
	\draw[thick, #6] (#1, #2) -- (#3, #4);
	\draw[thick, #6] (#3, #4) -- (#5, #4);
}

\newcommand{\strMinColor}[7]{
	\draw[thick, #7] (#1, #2) -- (#3, #2);
	\draw[thick, #7] (#3, #2) -- (#4, #5);
	\draw[thick, #7] (#4, #5) -- (#6, #5);
}

\newcommand{\circleColor}[3]{\draw[draw=#3, fill=#3](#1,#2) circle [radius=0.25cm]}

\newcommand{\fixed}[1]{#1^F}
\newcommand{\gv}[1]{#1}
\newcommand{\ongoingDom}{\Omega}
\newcommand{\boolDom}{\Gamma}

\newcommand{\ongoingBool}[3][]{
	\ifthenelse{\equal{#2}{\emptyset}}
	{\mathbf{b}[#2, \allowbreak \{#3\}]}
	{\ifthenelse{\equal{#3}{\emptyset}}
		{\mathbf{b}[\{#2\}, \allowbreak #3]}
			{\mathbf{b}[\{#2\}, \allowbreak \{#3\}]}
	}
}

\newcommand{\ongoingBoolVar}[2]{
	\mathbf{b}[#1, \allowbreak #2]
}

\newcommand{\ongoingBoolLetter}[1]{\mathbf{b}_{#1}}

\newcommand{\representOngoingBoolean}[1]{[#1]}

\newcommand{\rtattribute}[1]{\{#1\}}

\newcommand{\timeDom}{\mathcal{T}}
\newcommand{\minV}{{\textstyle \gv{\min}}}
\newcommand{\maxV}{{\textstyle \gv{\max}}}

\newcommand{\minF}{{\textstyle \fixed{\min}}}
\newcommand{\maxF}{{\textstyle \fixed{\max}}}

\newcommand{\smallplus}{\raisebox{1pt}{$\mathsmaller{\boldsymbol{+}}$}}

\newcommand{\Now}[1]{#1\smallplus}
\newcommand{\upNow}[1]{\smallplus#1}
\newcommand{\NowS}{now}
\newcommand{\minPoint}[2]{#1\smallplus#2}

\newcommand{\tcurr}{rt}
\def\twodig#1{\expandafter\ifnum#1<10 0\fi#1}
\newcommand{\dts}[2]{\twodig{#1}\textrm{/}\twodig{#2}}
\newcommand{\dtsYear}[2]{#1\textrm{/}\twodig{#2}}

\definecolor{ForestGreen}{RGB}{0,153,0}
\definecolor{DarkRed}{RGB}{204,0,0}

\newcommand{\ok}{\textcolor{ForestGreen}{\textbf{\checkmark}}}
\newcommand{\fail}{\textcolor{DarkRed}{\textbf{X}}}

\newcommand{\separateDisabled}{true}
\newcommand{\separate}{\ifthenelse{\equal{\separateDisabled}{true}}{}{\clearpage}}
\newcommand{\bind}[2]{\lVert #1 \rVert_{#2}}

\newcommand{\approachOngoingRelations}{
  \begin{figure}[!htb] \centering
    \begin{tikzpicture}[shorten >=1pt, auto, node distance = 3cm,
      thick,scale=1, every node/.style={scale=0.77}]

      \def\xa{0}
      \def\ya{0}

		\draw(\xa,\ya) node[anchor=north,align=left] {
            \begin{tabular}[t]{>{\raggedleft\arraybackslash}p{.2cm}|ccll|}
			\multicolumn{5}{l}{\hspace{15pt}$\mathbf{B}$} \\ \cline{2-5}
			& BID & C   & {\hspace{15pt} VT} & {\hspace{15pt} RT}\\ \cline{2-5}
			$b_1$ & 500 & Spam filter & $[\dts{1}{25}, \NowS)$ & $\rtattribute{(-\infty, \infty)}$\\
			$b_2$ & 501 & Spam filter & $[\dts{3}{30}, \dts{8}{21})$ & $\rtattribute{(-\infty, \infty)}$\\
			\cline{2-5}
			\end{tabular}
            \\[7pt]
			\begin{tabular}[t]{>{\raggedleft\arraybackslash}p{.2cm}|ccll|}
			\multicolumn{5}{l}{\hspace{15pt}$\mathbf{P}$} \\ \cline{2-5}
			& PID & C & {\hspace{15pt} VT} & {\hspace{15pt} RT} \\ \cline{2-5}
			$p_1$ & 201 & Spam filter & $[\dts{8}{15}, \dts{8}{24})$ & $\rtattribute{(-\infty, \infty)}$ \\
			$p_2$ & 202 & Spam filter & $[\dts{8}{24}, \dts{8}{27})$ & $\rtattribute{(-\infty, \infty)}$ \\
			\cline{2-5}
			\end{tabular}
			\\[7pt]
			\begin{tabular}[t]{>{\raggedleft\arraybackslash}p{.2cm}|ccll|}
			\multicolumn{5}{l}{\hspace{15pt}$\mathbf{L}$} \\ \cline{2-5}
			& Name & C & {\hspace{15pt} VT} & {\hspace{15pt} RT} \\ \cline{2-5}
			$l_1$ & Ann & Spam filter & $[\dts{1}{20}, \dts{8}{18})$ & $\rtattribute{(-\infty, \infty)}$ \\
			$l_2$ & Bob & Spam filter & $[\dts{8}{18}, \NowS)$ & $\rtattribute{(-\infty, \infty)}$ \\
			\cline{2-5}
			\end{tabular}
		};
		\end{tikzpicture}
		\caption{Relations with ongoing time points.}
		\label{fig:exampleOngoingRelations}
	\end{figure}
}

\newcommand{\approachOverviewBeforeQuery}{
  \begin{figure}[!htb] \centering \setlength{\tabcolsep}{3pt}
    \begin{tikzpicture}[shorten >=1pt, auto, node distance = 3cm,
      thick,scale=1, every node/.style={scale=0.77}]

		\def\xa{0}
		\def\yb{-0.8}
		\def\yc{0}

		\draw (\xa,\yb) node[anchor=north,align=center] {
			\begin{tabular}[t]{l@{\hspace{2pt}}|clccll|}
			\multicolumn{5}{l}{\hspace{15pt}$\mathbf{V}$} \\ \cline{2-7}
			& BID & {\hspace{15pt} $\mathbf{B}$.VT} & PID & Name & {\hspace{10pt} $\mathbf{B}.\text{VT} \gv{\cap} \mathbf{L}.\text{VT}$} & {\hspace{15pt} RT}\\ \cline{2-7}
			$v_1$ & 500 & $[\dts{1}{25}, \NowS)$ & 201 & Ann & $[\dts{1}{25}, \upNow{\dts{8}{18}})$ & $\rtattribute{[\dts{1}{26}, \dts{8}{16})}$ \\
			$v_2$ & 500 & $[\dts{1}{25}, \NowS)$ & 202 & Ann & $[\dts{1}{25}, \upNow{\dts{8}{18}})$ & $\rtattribute{[\dts{1}{26}, \dts{8}{25})}$ \\
			$v_3$ & 500 & $[\dts{1}{25}, \NowS)$ & 202 & Bob & $[\dts{8}{18}, \NowS)$ & $\rtattribute{[\dts{8}{19}, \dts{8}{25})}$ \\
			$v_4$ & 501 & $[\dts{3}{30}, \dts{8}{21})$ & 202 & Ann & $[\dts{3}{30}, \dts{8}{18})$ & $\rtattribute{(-\infty, \infty)}$ \\
			$v_5$ & 501 & $[\dts{3}{30}, \dts{8}{21})$ & 202 & Bob & $[\dts{8}{18}, \upNow{\dts{8}{21}})$ & $\rtattribute{[\dts{8}{19}, \infty)}$ \\
			\cline{2-7}
			\end{tabular}
		};
		\end{tikzpicture}
		\caption{Query result $\mathbf{V}$ remains valid as time passes by.}
		\label{fig:exampleMatViewBefore}
	\end{figure}
}

\newcommand{\nonCoreOps}{
	\begin{table}[!htb]
		\fontsize{7}{9}\selectfont
		\caption{Equivalences for predicates and function on ongoing time points and time intervals.}
		\renewcommand{\arraystretch}{1.6}
		\setlength\arraycolsep{3pt}
		\begin{centermath}
			\begin{array}{cl} \hline \textbf{Operation}
				& \textbf{Equivalence} \\ \hline
				\defcolorrow
				\gv{\leq}
				& t_1 \gv{\leq} t_2 \equiv \gv{\neg}(t_2 \gv{<} t_1)
				\\
				& \multicolumn{1}{l}{\examplecolorrow
					\textbf{Example} \hspace*{0.25cm}
					\begin{aligned}[t]
						&\NowS \gv{\leq} \dts{10}{17} =
						\gv{\neg}(\ongoingBool{[\dts{10}{18}, \infty)}{(-\infty, \dts{10}{18})}) \\
						&= \ongoingBool{(-\infty, \dts{10}{18})}{[\dts{10}{18}, \infty)}
				\end{aligned}}
				\\
				\defcolorrow
				\gv{=}
				& t_1 \gv{=} t_2 \equiv t_1 \gv{\leq} t_2 \gv{\wedge} t_2 \gv{\leq} t_1
				\\
				& \multicolumn{1}{l}{\examplecolorrow
					\textbf{Example} \hspace*{0.25cm}
					\begin{aligned}[t]
						&(\dts{10}{17} \gv{=} \NowS) \\
						&= \ongoingBool[r]{[\dts{10}{17}, \dts{10}{18})}
						{(-\infty, \dts{10}{17}), [\dts{10}{18}, \infty)}
					\end{aligned}
				}
				\crc{\\
				\defcolorrow
				\gv{\neq}
				& t_1 \gv{\neq} t_2 \equiv (t_1 \gv{<} t_2) \gv{\vee} (t_2 \gv{<} t_1)
				\\
				& \multicolumn{1}{l}{\examplecolorrow
					\textbf{Example} \hspace*{0.25cm}
					\begin{aligned}[t]
						&\dts{10}{17} \gv{\neq} \NowS \\
						&= \ongoingBool[r]{(-\infty, \dts{10}{17}), [\dts{10}{18}, \infty)}{[\dts{10}{17}, \dts{10}{18})}
					\end{aligned}
				}}
				\\ \hline
				\defcolorrow
				\gv{\text{before}}
				& [t_s, t_e)\ \gv{\text{before}}\ [\tilde{t}_s, \tilde{t}_e) \equiv
				t_e \gv{\leq} \tilde{t}_s \gv{\wedge} t_s \gv{<} t_e
				\gv{\wedge} \tilde{t}_s \gv{<} \tilde{t}_e
				\\
				& \multicolumn{1}{l}{\examplecolorrow
					\textbf{Example} \hspace*{0.25cm}
					\begin{aligned}[t]
						& [\dts{10}{17}, \NowS) \ \gv{\text{before}} \ [\dts{10}{20}, \dts{10}{25}) \\
						& = \ongoingBool[r]{[\dts{10}{18}, \dts{10}{21})}{(-\infty, \dts{10}{18}), [\dts{10}{21}, \infty)}
					\end{aligned}

				}
				\\
				\crc{
				\defcolorrow
				\gv{\text{meets}}
				& [t_s, t_e)\ \gv{\text{meets}}\ [\tilde{t}_s, \tilde{t}_e) \equiv
				t_e \gv{=} \tilde{t}_s \gv{\wedge} t_s \gv{<} t_e
				\gv{\wedge} \tilde{t}_s \gv{<} \tilde{t}_e
				\\
				& \multicolumn{1}{l}{\examplecolorrow
					\textbf{Example} \hspace*{0.25cm}
					\begin{aligned}[t]
						&[\dts{10}{17}, \NowS) \ \gv{\text{meets}} \ [\dts{10}{20}, \dts{10}{25}) \\
						&	= \ongoingBool[r]{[\dts{10}{20}, \dts{10}{21})}
						{(-\infty, \dts{10}{20}), [\dts{10}{21}, \infty)}
					\end{aligned}

				}
				\\
				}
				\defcolorrow
				\gv{\text{overlaps}}
				& [t_s, t_e)\ \gv{\text{overlaps}}\ [\tilde{t}_s, \tilde{t}_e) \equiv
				t_s \gv{<} \tilde{t}_e \gv{\wedge} \tilde{t}_s \gv{<} t_e
				\gv{\wedge} t_s \gv{<} t_e \gv{\wedge} \tilde{t}_s \gv{<} \tilde{t}_e
				\\
				& \multicolumn{1}{l}{\examplecolorrow
					\textbf{Example} \hspace*{0.25cm}
					\begin{aligned}[t]
						&[\dts{10}{17}, \NowS) \ \gv{\text{overlaps}} \ [\dts{10}{14}, \dts{10}{20}) \\
						&= \ongoingBool[r]{[\dts{10}{18}, \infty)}{(-\infty, \dts{10}{18})}
					\end{aligned}
				}\\
				\crc{
				\defcolorrow
				\gv{\text{starts}}
				& [t_s, t_e)\ \gv{\text{starts}}\ [\tilde{t}_s, \tilde{t}_e) \equiv
				t_s \gv{=} \tilde{t}_s \gv{\wedge} t_s \gv{<} t_e
				\gv{\wedge} \tilde{t}_s \gv{<} \tilde{t}_e
				\\
				& \multicolumn{1}{l}{\examplecolorrow
					\textbf{Example} \hspace*{0.25cm}
					\begin{aligned}[t]
						&[\dts{10}{17}, \NowS) \ \gv{\text{starts}} \ [\dts{10}{17}, \dts{10}{20}) \\
						&= \ongoingBool[r]{[\dts{10}{18}, \infty)}{(-\infty, \dts{10}{18})}
					\end{aligned}
				}
				\\}
			\crc{
				\defcolorrow
				\gv{\text{finishes}}
				& [t_s, t_e)\ \gv{\text{finishes}}\ [\tilde{t}_s, \tilde{t}_e) \equiv
				t_e \gv{=} \tilde{t}_e \gv{\wedge} t_s \gv{<} t_e
				\gv{\wedge} \tilde{t}_s \gv{<} \tilde{t}_e
				\\
				& \multicolumn{1}{l}{\examplecolorrow
					\textbf{Example} \hspace*{0.25cm}
					\begin{aligned}[t]
						&[\dts{10}{17}, \NowS) \ \gv{\text{finishes}} \ [\dts{10}{20}, \dts{10}{25}) \\
						&= \ongoingBool[r]{[\dts{10}{25}, \dts{10}{26})}
						{(-\infty, \dts{10}{25}), [\dts{10}{26}, \infty)}
					\end{aligned}
				}
			\\
			}

			\crc{
				\defcolorrow
				\gv{\text{during}}
				& \begin{aligned}[t]
					[t_s, t_e)\ \gv{\text{during}}\ [\tilde{t}_s, \tilde{t}_e)  \equiv\
					& (\tilde{t}_s \gv{\leq} t_s \gv{\wedge} t_e \gv{\leq} \tilde{t}_e
					\gv{\wedge} t_s \gv{<} t_e \gv{\wedge} \tilde{t}_s \gv{<} \tilde{t}_e) \\
					&\gv{\vee} (t_e \gv{\leq} t_s \gv{\wedge} \tilde{t}_s \gv{<} \tilde{t}_e)
				\end{aligned}
				\\
				& \multicolumn{1}{l}{\examplecolorrow
					\textbf{Example} \hspace*{0.25cm}
					\begin{aligned}[t]
						&[\dts{10}{20}, \dts{10}{25}) \ \gv{\text{during}} \ [\dts{10}{17}, \NowS) \\
						&= \ongoingBool[r]{[\dts{10}{25}, \infty)}{(-\infty, \dts{10}{25})}
					\end{aligned}
				}
				\\
			}
				\crc{
				\defcolorrow
				\gv{\text{equals}}
				& \begin{aligned}[t]
					[t_s, t_e)\ \gv{\text{equals}}\ [\tilde{t}_s, \tilde{t}_e) \equiv\
					&(t_s \gv{=} \tilde{t}_s \gv{\wedge} t_e \gv{=} \tilde{t}_e
					\gv{\wedge} t_s \gv{<} t_e \gv{\wedge} \tilde{t}_s \gv{<} \tilde{t}_e) \\
					&\gv{\vee} (t_e \gv{\leq} t_s \gv{\wedge} \tilde{t}_e \gv{\leq} \tilde{t}_s)
				\end{aligned}
				\\
				& \multicolumn{1}{l}{\examplecolorrow
					\textbf{Example} \hspace*{0.25cm}
					\begin{aligned}[t]
						&[\dts{10}{17}, \NowS) \ \gv{\text{equals}} \ [\dts{10}{17}, \dts{10}{20}) \\
						&= \ongoingBool[r]{[\dts{10}{20}, \dts{10}{21})}
						{(-\infty, \dts{10}{20}), [\dts{10}{21}, \infty)}
					\end{aligned}

				}
				\\
			}
			 \hline
				\defcolorrow
				\gv{\cap}
				& [t_s, t_e) \gv{\cap} [\tilde{t}_s, \tilde{t}_e) \equiv
				[\maxV(t_s, \tilde{t}_s), \minV(t_e, \tilde{t}_e))
				\\
				& \multicolumn{1}{l}{\examplecolorrow
					\textbf{Example} \hspace*{0.25cm}
					\begin{aligned}[t]
						&[\dts{10}{17}, \NowS) \gv{\cap} [\dts{10}{14}, \dts{10}{20}) = [\dts{10}{17}, \upNow{\dts{10}{20}})
					\end{aligned}
				}
				\\ \hline
			\end{array}
		\end{centermath}
		\label{fig:remainingOps}
		\vspace*{-9pt}
	\end{table}
}

\newcommand{\tableOngoingTimePoints}{
\begin{figure*}[!ht]
	\renewcommand{\arraystretch}{1.3}
	\footnotesize
	\centering
	\begin{tabular}[t]{lccccc}
		\hline
		& \multirow{2}{*}{\centering\textbf{Ongoing time point}}
		& \multicolumn{4}{c}{\textbf{Type}}
		\\ \cline{3-6}
		&
		& \textbf{Fixed time point} & \textbf{Time point $\NowS$}
		& \textbf{Growing time point} & \textbf{Limited time point}
		\\ \hline
		Notation & & & & & \\
		- as $\minPoint{a}{b}$ & $\minPoint{a}{b}$ &
		$\minPoint{a}{a}$
		& $\minPoint{-\infty}{\infty}$ & $\minPoint{a}{\infty}$
		& $\minPoint{-\infty}{b}$ \\
		- short &
		$\minPoint{a}{b}$ & $a$ & $\NowS$ & $\Now{a}$ & $\upNow{b}$ \\
		Meaning & not earlier than $a$, & time point $a$ & the \emph{current} time point & not earlier than $a$, & possibly earlier, \\[-4pt]
		& \hspace{0.1cm} but not later than $b$ & & & \hspace{-0.5cm} possibly later & \hspace{0.3cm} but not later than $b$ \\
		Example &
		$\minPoint{\dts{10}{17}}{\dts{10}{19}}$
		&
		$\dts{10}{17}$ & $\NowS$
		& $\Now{\dts{10}{17}}$ & $\upNow{\dts{10}{17}}$ \\
		Semantics &
		\adjustbox{valign=t}{
			\begin{tikzpicture}[anchor=north,scale = 0.17,every node/.style={scale=0.8}]
			\coordSystem{7}{7};

			\strMinColor{0}{2.5}{2.5}{4.5}{4.5}{7}{white!30!black};
			\circleColor{0.5}{2.5}{dkgreen};
			\circleColor{1.5}{2.5}{dkgreen};
			\circleColor{2.5}{2.5}{dkgreen};
			\circleColor{3.5}{3.5}{dkgreen};
			\circleColor{4.5}{4.5}{dkgreen};
			\circleColor{5.5}{4.5}{dkgreen};
			\circleColor{6.5}{4.5}{dkgreen};

			\draw (-0.3,2.5) node[left] {$\dts{10}{17}$} -- (0.3,2.5);
			\draw (-0.3,4.5) node[left] {$\dts{10}{19}$} -- (0.3,4.5);
			\draw (2.5,-0.3) node[below] {$\dts{10}{17}$} -- (2.5,0.3);
			\end{tikzpicture}
		}\hspace*{-0.3cm}
		&
		\adjustbox{valign=t}{
			\begin{tikzpicture}[anchor=north,scale = 0.17,every node/.style={scale=0.8}]
			\coordSystem{7}{7};
			\strLineColor{0}{2.5}{7}{white!30!black};
			\circleColor{0.5}{2.5}{darkred};
			\circleColor{1.5}{2.5}{darkred};
			\circleColor{2.5}{2.5}{darkred};
			\circleColor{3.5}{2.5}{darkred};
			\circleColor{4.5}{2.5}{darkred};
			\circleColor{5.5}{2.5}{darkred};
			\circleColor{6.5}{2.5}{darkred};
			\draw (-0.3,2.5) node[left] {$\dts{10}{17}$} -- (.3,2.5);
			\draw (2.5,-0.3) node[below] {$\dts{10}{17}$} -- (2.5,0.3);
			\end{tikzpicture}
		}\hspace*{-0.3cm}
		&
		\adjustbox{valign=t}{
			\begin{tikzpicture}[anchor=north,scale = 0.17,every node/.style={scale=0.8}]
			\coordSystem{7}{7};
			\strNowColor{0}{0}{0}{7}{7}{white!30!black};
			\circleColor{0.5}{0.5}{orangeDark};
			\circleColor{1.5}{1.5}{orangeDark};
			\circleColor{2.5}{2.5}{orangeDark};
			\circleColor{3.5}{3.5}{orangeDark};
			\circleColor{4.5}{4.5}{orangeDark};
			\circleColor{5.5}{5.5}{orangeDark};
			\circleColor{6.5}{6.5}{orangeDark};

			\draw (-0.3,2.5) node[left] {$\dts{10}{17}$} -- (0.3,2.5);
			\draw (2.5,-0.3) node[below] {$\dts{10}{17}$} -- (2.5,0.3);
			\end{tikzpicture}
		}\hspace*{-0.3cm}
		&
		\adjustbox{valign=t}{
			\begin{tikzpicture}[anchor=north,scale = 0.17,every node/.style={scale=0.8}]
			\coordSystem{7}{7};
			\strNowColor{0}{2.5}{2.5}{7}{7}{white!30!black};
			\circleColor{0.5}{2.5}{lightBlue};
			\circleColor{1.5}{2.5}{lightBlue};
			\circleColor{2.5}{2.5}{lightBlue};
			\circleColor{3.5}{3.5}{lightBlue};
			\circleColor{4.5}{4.5}{lightBlue};
			\circleColor{5.5}{5.5}{lightBlue};
			\circleColor{6.5}{6.5}{lightBlue};

			\draw (-0.3,2.5) node[left] {$\dts{10}{17}$} -- (.3,2.5);
			\draw (2.5,-0.3) node[below] {$\dts{10}{17}$} -- (2.5,0.3);
			\end{tikzpicture}
		}\hspace*{-0.3cm}
		&
		\adjustbox{valign=t}{
			\begin{tikzpicture}[anchor=north,scale = 0.17,every node/.style={scale=0.8}]
			\coordSystem{7}{7};
			\strUpNowColor{0}{0}{2.5}{2.5}{7}{white!30!black};
			\circleColor{0.5}{0.5}{mauve};
			\circleColor{1.5}{1.5}{mauve};
			\circleColor{2.5}{2.5}{mauve};
			\circleColor{3.5}{2.5}{mauve};
			\circleColor{4.5}{2.5}{mauve};
			\circleColor{5.5}{2.5}{mauve};
			\circleColor{6.5}{2.5}{mauve};

			\draw (-0.3,2.5) node[left] {$\dts{10}{17}$} -- (0.3,2.5);
			\draw (2.5,-0.3) node[below] {$\dts{10}{17}$} -- (2.5,0.3);
			\end{tikzpicture}
		}
		\\
		\hline
	\end{tabular}
	\caption{Illustration of ongoing time points $\minPoint{a}{b}$.}
	\label{fig:timeDomain}
\end{figure*}
}

\newcommand{\tableOngoingIntervals}{
\begin{figure*}[!bhtp]
	\centering
	\footnotesize
	\renewcommand{\arraystretch}{1.3}
	\begin{tabular}[t]{lcccc}
		\hline
		&\centering \multirow{2}{*}{\textbf{Ongoing time interval}} & \multicolumn{3}{c}{\textbf{Type}} \\ \cline{3-5}
		& & \textbf{Fixed time interval} & \textbf{Expanding time interval} & \textbf{Shrinking time interval} \\ \hline
		\defcolorrow
		\emph{\textbf{non-empty}} &
		\centering if $a \leq b < c \leq d$ & if $a = b < c = d$ & if $a = b < c < d$ & if $a < b < c = d$   \\
		& $[\minPoint{\dts{10}{16}}{\dts{10}{17}},\minPoint{\dts{10}{19}}{\dts{10}{20}})$
		& $[\dts{10}{17},\dts{10}{19})$
		& $[\dts{10}{17},\minPoint{\dts{10}{19}}{\dts{10}{21}})$
		& $[\upNow{\dts{10}{17}},\dts{10}{19})$ \\
		& \adjustbox{valign=t}{
			\begin{tikzpicture}[anchor=north,scale = 0.17,every node/.style={scale=0.8}]
			\coordSystem{8}{8};

			\strMinColor{0}{1.5}{1.5}{2.5}{2.5}{7}{black};
			\strMinColor{0}{4.5}{4.5}{5.5}{5.5}{7}{black};

			\strLineColorY{0.5}{1.5}{4.5}{dkgreen};
			\strLineColorY{1.5}{1.5}{4.5}{dkgreen};
			\strLineColorY{2.5}{2.5}{4.5}{dkgreen};
			\strLineColorY{3.5}{2.5}{4.5}{dkgreen};
			\strLineColorY{4.5}{2.5}{4.5}{dkgreen};
			\strLineColorY{5.5}{2.5}{5.5}{dkgreen};
			\strLineColorY{6.5}{2.5}{5.5}{dkgreen};

			\draw (-0.3,2.5) node[left] {$\dts{10}{17}$} -- (0.3,2.5);
			\draw (-0.3,4.5) node[left] {$\dts{10}{19}$} -- (0.3,4.5);
			\draw (-0.3,1.5) -- (0.3,1.5);
			\draw (2.5,-0.3) node[below] {$\dts{10}{17}$} -- (2.5,0.3);
			\end{tikzpicture}
		}
		& \adjustbox{valign=t}{
			\begin{tikzpicture}[anchor=north,scale = 0.17,every node/.style={scale=0.8}]
			\coordSystem{8}{8};

			\strLineColor{0}{2.5}{8}{black};
			\strLineColor{0}{4.5}{8}{black};

			\strLineColorY{0.5}{2.5}{4.5}{darkred};
			\strLineColorY{1.5}{2.5}{4.5}{darkred};
			\strLineColorY{2.5}{2.5}{4.5}{darkred};
			\strLineColorY{3.5}{2.5}{4.5}{darkred};
			\strLineColorY{4.5}{2.5}{4.5}{darkred};
			\strLineColorY{5.5}{2.5}{4.5}{darkred};
			\strLineColorY{6.5}{2.5}{4.5}{darkred};
			\strLineColorY{7.5}{2.5}{4.5}{darkred};

			\draw (-0.3,2.5) node[left] {$\dts{10}{17}$} -- (.3,2.5);
			\draw (-0.3,4.5) node[left] {$\dts{10}{19}$} -- (.3,4.5);
			\draw (2.5,-0.3) node[below] {$\dts{10}{17}$} -- (2.5,0.3);
			\end{tikzpicture}
		}
		&
		\adjustbox{valign=t}{
			\begin{tikzpicture}[anchor=north,scale = 0.17,every node/.style={scale=0.8}]
			\coordSystem{8}{8};

			\strLineColor{0}{2.5}{8}{black};
			\strMinColor{0}{4.5}{4.5}{6.5}{6.5}{8}{black};
			\strLineColorY{0.5}{2.5}{4.5}{lightBlue};
			\strLineColorY{1.5}{2.5}{4.5}{lightBlue};
			\strLineColorY{2.5}{2.5}{4.5}{lightBlue};
			\strLineColorY{3.5}{2.5}{4.5}{lightBlue};
			\strLineColorY{4.5}{2.5}{4.5}{lightBlue};
			\strLineColorY{5.5}{2.5}{5.5}{lightBlue};
			\strLineColorY{6.5}{2.5}{6.5}{lightBlue};
			\strLineColorY{7.5}{2.5}{6.5}{lightBlue};

			\draw (-0.3,2.5) node[left] {$\dts{10}{17}$} -- (.3,2.5);
			\draw (-0.3,4.5) node[left] {$\dts{10}{19}$} -- (.3,4.5);
			\draw (2.5,-0.3) node[below] {$\dts{10}{17}$} -- (2.5,0.3);
			\end{tikzpicture}
		}

		&
		\adjustbox{valign=t}{
			\begin{tikzpicture}[anchor=north,scale = 0.17,every node/.style={scale=0.8}]
			\coordSystem{8}{8};

			\strLineColor{0}{4.5}{8}{black};
			\strUpNowColor{0}{0}{2.5}{2.5}{8}{black};
			\strLineColorY{0.5}{0.5}{4.5}{mauve};
			\strLineColorY{1.5}{1.5}{4.5}{mauve};
			\strLineColorY{2.5}{2.5}{4.5}{mauve};
			\strLineColorY{3.5}{2.5}{4.5}{mauve};
			\strLineColorY{4.5}{2.5}{4.5}{mauve};
			\strLineColorY{5.5}{2.5}{4.5}{mauve};
			\strLineColorY{6.5}{2.5}{4.5}{mauve};
			\strLineColorY{7.5}{2.5}{4.5}{mauve};

			\draw (-0.3,2.5) node[left] {$\dts{10}{17}$} -- (0.3,2.5);
			\draw (-0.3,4.5) node[left] {$\dts{10}{19}$} -- (.3, 4.5);
			\draw (2.5,-0.3) node[below] {$\dts{10}{17}$} -- (2.5,0.3);
			\end{tikzpicture}
		}
		\\ \hline
		\defcolorrow
		\emph{\textbf{partially empty}} &
		\centering if $a < c \leq b$ or $c \leq a \leq b < d$ & never & if $c \leq a \leq b < d$ & if $a < c \leq d \leq b$ \\
		& $[\minPoint{\dts{10}{16}}{\dts{10}{19}},\minPoint{\dts{10}{17}}{\dts{10}{20}})$& -- & $[\dts{10}{17}, \NowS)$ & $[\Now{\dts{10}{16}}, \dts{10}{19})$ \\
		& \adjustbox{valign=t}{
			\begin{tikzpicture}[anchor=north,scale = 0.17,every node/.style={scale=0.8}]
			\coordSystem{7}{8};

			\strMinColor{0}{1.5}{1.5}{4.5}{4.5}{7}{black};
			\strMinColor{0}{2.5}{2.5}{5.5}{5.5}{7}{black};

			\strLineColorY{0.5}{1.5}{2.5}{dkgreen};
			\strLineColorY{1.5}{1.5}{2.5}{dkgreen};
			\strLineColorY{5.5}{4.5}{5.5}{dkgreen};
			\strLineColorY{6.5}{4.5}{5.5}{dkgreen};

			\draw (-0.3,2.5) node[left] {$\dts{10}{17}$} -- (0.3,2.5);
			\draw (-0.3,4.5) node[left] {$\dts{10}{19}$} -- (0.3,4.5);
			\draw (2.5,-0.3) node[below] {$\dts{10}{17}$} -- (2.5,0.3);
			\end{tikzpicture}
		}
		& &
		\adjustbox{valign=t}{
			\begin{tikzpicture}[anchor=north,scale = 0.17,every node/.style={scale=0.8}]
			\coordSystem{9}{8};

			\strLineColor{2.5}{2.5}{8}{black};
			\strNowColor{2.5}{2.5}{2.5}{8}{8}{black};
			\strLineColorY{3.5}{2.5}{3.5}{lightBlue};
			\strLineColorY{4.5}{2.5}{4.5}{lightBlue};
			\strLineColorY{5.5}{2.5}{5.5}{lightBlue};
			\strLineColorY{6.5}{2.5}{6.5}{lightBlue};
			\strLineColorY{7.5}{2.5}{7.5}{lightBlue};

			\draw (-0.3,2.5) node[left] {$\dts{10}{17}$} -- (0.3,2.5);
			\draw (2.5,-0.3) node[below] {$\dts{10}{17}$} -- (2.5,0.3);
			\end{tikzpicture}
		}
		&
		\adjustbox{valign=t}{
			\begin{tikzpicture}[anchor=north,scale = 0.17,every node/.style={scale=0.8}]
			\coordSystem{9}{8};

			\strLineColor{0}{4.5}{4.5}{black};
			\strMinColor{0}{1.5}{1.5}{4.5}{4.5}{4.5}{black};
			\strLineColorY{0.5}{1.5}{4.5}{mauve};
			\strLineColorY{1.5}{1.5}{4.5}{mauve};
			\strLineColorY{2.5}{2.5}{4.5}{mauve};
			\strLineColorY{3.5}{3.5}{4.5}{mauve};

			\draw (-0.3,1.5) node[left] {$\dts{10}{16}$} -- (.3, 1.5);
			\draw (-0.3,4.5) node[left] {$\dts{10}{19}$} -- (.3, 4.5);
			\draw (4.5,-0.3) node[below] {$\dts{10}{19}$} -- (4.5,0.3);
			\end{tikzpicture}
		}
		\\ \hline
	\end{tabular}
	\caption{Illustration of ongoing time intervals $[\minPoint{a}{b}, \minPoint{c}{d})$.}
	\label{fig:ongoingTimeIntervals}
\end{figure*}
}

\newcommand{\examplecolorrow}{}
\newcommand{\defcolorrow}{\rowcolor{gray!20}}

\newcommand{\thetaOverlaps}{\text{ovlp}}
\newcommand{\thetaBefore}{\text{bef}}

\AtBeginDocument{
	\mathchardef\mathcomma\mathcode`\,
	\mathcode`\,="8000
}
{\catcode`,=\active
	\gdef,{\mathcomma\discretionary{}{}{}}
}

\newtheorem{lemma}{Theorem}
\newtheorem{definition}{Definition}
\newtheorem{example}{Example}

\crefname{figure}{Fig.}{Fig.}
\crefname{lemma}{Theorem}{Theorem}

\newcommand\blfootnote[1]{%
	\begingroup
	\renewcommand\thefootnote{}\footnote{#1}%
	\addtocounter{footnote}{-1}%
	\endgroup
}

\begin{document}

\title{Query Results over Ongoing Databases that Remain Valid as Time Passes By \\ (Extended Version*)
}

\author{\IEEEauthorblockN{Yvonne M\"{u}lle}
	\IEEEauthorblockA{\textit{Department of Computer Science} \\
		\textit{University of Zurich}\\
		Switzerland\\
		Email: muelle@ifi.uzh.ch}
	\and
	\IEEEauthorblockN{Michael H. B\"ohlen}
	\IEEEauthorblockA{\textit{Department of Computer Science} \\
		\textit{University of Zurich}\\
		Switzerland\\
		Email: boehlen@ifi.uzh.ch}
}

\allowdisplaybreaks[1]

\hyphenation{ mi-ni-mum ma-xi-mum align-ment da-ta-ba-ses da-ta-base
  ca-non-i-cal-i-za-tion time-stamp time-stamps wheth-er over-view
  pre-sent bool-eans bool-ean emp-ty que-ry MozillaBugs on-go-ing }

\maketitle

\blfootnote{$^*$ Extended version of the ICDE paper \cite{generallyValidQueries}}

\begin{abstract}
  Ongoing time point $\NowS$ is used to state that a tuple is valid
  from the start point onward. For database systems ongoing time
  points have far-reaching implications since they change continuously
  as time passes by. State-of-the-art approaches deal with ongoing
  time points by instantiating them to the reference time. The
  instantiation yields query results that are only valid at the chosen
  time and get invalidated as time passes by.

  We propose a solution that keeps ongoing time points
  \emph{uninstantiated} during query processing. We do so by
  evaluating predicates and functions at all possible reference
  times. This renders query results independent of a specific
  reference time and yields results that remain valid as time passes
  by. As query results, we propose \emph{ongoing relations} that
  include a \emph{reference time attribute}. The value of the
  reference time attribute is restricted by predicates and functions
  on ongoing attributes. We describe and evaluate an efficient
  implementation of ongoing data types and operations in PostgreSQL.
\end{abstract}

\section{Introduction}
\label{sec:introduction}

Data that are associated with a valid time interval \cite{validTime}
are present in real-world applications that deal with employment
contracts, insurance policies, software bugs, etc. The ongoing time
point $\NowS$ is commonly used to state that the contract, policy,
bug, etc. is valid from the start point onward.

The ongoing time point $\NowS$ changes its value when time passes by
and the reference time is used to determine the value.  At each
reference time, $\NowS$ instantiates to the time point equal to the
reference time.  For example, at reference time $\dts{8}{15}$, $\NowS$
instantiates to time point $\dts{8}{15}$ and at reference time
$\dts{8}{16}$, it instantiates to time point $\dts{8}{16}$. Throughout
the paper, we use time points in the $\text{mm}/\text{dd}$ format
relative to 2019: time point $\dts{8}{15}$ denotes August 15, 2019.

A key assumption of database systems is that query
results only get outdated if data is modified explicitly.  This
happens if data is inserted, updated, or deleted.  The assumption no
longer holds if $\NowS$ is stored in the database or when queries
are evaluated on databases with ongoing time points
\cite{now,modificationSemanaticsNow, nowQuerying}.  In this case,
query results get also outdated as a result of time passing by.  This has
significant drawbacks.  First, query results, including materialized
views and cached query results, must be re-computed before they can
be accessed.  Second, because ongoing time points are replaced by
fixed time points, it is impossible for applications to identify
result time points that change when time passes by.

This paper proposes an elegant and efficient solution that preserves
ongoing time points in query results and that evaluates queries at all
possible reference times to get results that remain valid as time
passes by. Formally, given a database $D$ with ongoing time
points and a query $Q$, we want to compute a query result $Q(D)$,
such that at every possible reference time $\tcurr$, the query
result is equivalent to the result obtained by instantiating $\NowS$
in $D$ and evaluating the query on the instantiated database:
$\forall \tcurr (\bind{Q(D)}{\tcurr} \equiv Q(\bind{D}{\tcurr}))$.
The bind operator $\bind{\cdot}{\tcurr}$ replaces all occurrences of
$\NowS$ with the reference time $\tcurr$.

To support queries with predicates and functions on ongoing
attributes, the key challenges are
\begin{enumerate*}[label=(\arabic*)]
\item the evaluation of queries to results that remain valid as time
  passes by and
\item the representation of these results.
\end{enumerate*}

To get results that remain valid, we keep ongoing time points
uninstantiated. We define six core operations predicate $\gv{<}$,
functions $\minV$ and $\maxV$, and the logical connectives
$\gv{\wedge}$, $\gv{\vee}$, $\gv{\neg}$. At each reference time, their
results are equal to the results obtained by the corresponding
operations for fixed data types on the instantiated input arguments.
We provide equivalences for the core operations and for additional
operations that are expressed with the core operations. The
equivalences are used for an efficient implementation.  We represent
the results of predicates and logical connectives as \emph{ongoing
  booleans}, i.e., booleans whose truth value depends on the time.
The results of relational algebra operators are represented as
\emph{ongoing relations} that include a reference time attribute
$RT$. The value of $RT$ includes the reference times when a tuple
belongs to the instantiated relations.  The reference time of a tuple
is restricted by predicates in queries.  We represent the value of the
$RT$ attribute with a finite set of fixed time intervals.  Thus, only
predicates that evaluate to booleans that change their value a finite
number of times are allowed.  The tuples in base ongoing relations
have a trivial reference time, i.e.,
$RT = \rtattribute{(-\infty, \infty)}$. Tuples with an empty reference
time, i.e., $RT = \rtattribute{}$, are deleted.

Our technical contributions are the following:
\begin{itemize}
\item We propose the ongoing time domain $\ongoingDom$ for
  \emph{ongoing time points}.  The time domain is closed for $\minV$
  and $\maxV$, i.e., the evaluation of $\minV$ and $\maxV$ on
  $\ongoingDom$ again yields an ongoing time point of $\ongoingDom$.
\item We define predicates, functions and logical connectives that
  keep ongoing time points uninstantiated during query processing.
\item We introduce \emph{ongoing relations} with a reference time
  attribute to represent query results that remain valid as time
  passes by. The value of the $RT$ attribute is set by the database
  system and restricted by predicates on ongoing attributes.
\item We define the relational algebra for ongoing relations.  The
  result of each operator is an ongoing relation that remains valid as
  time passes by.
\item We describe an efficient implementation of ongoing data types
  and operations on these data types in the kernel of PostgreSQL.
\end{itemize}

The paper is organized as follows. \Cref{ex:applicationExample}
introduces our running example. \Cref{sec:relWork} discusses related
work. \Cref{sec:background} provides preliminaries.  We define the
time domain for ongoing time points in \Cref{sec:ongoingDataTypes}.
Predicates and functions on ongoing time points and time
  intervals whose results remain valid are discussed in
\Cref{sec:gvOperations}.  \Cref{sec:operators} introduces ongoing
relations and defines a relational algebra on them.
\Cref{sec:implementation} discusses the implementation of our solution
in PostgreSQL.  The evaluation is described in \Cref{sec:evaluation}.
\Cref{sec:conclusion} concludes the paper and points to future
research.

\section{Running Example}
\label{ex:applicationExample}

Consider a company that keeps track of bugs associated with the
individual components of its email service.  Prioritized bugs have
fixed start points that indicate when the bug was discovered and fixed
end points that indicate the deadline for resolving the bug
internally.  Deprioritized bugs have fixed start points but end points
that keep increasing. These end points are \emph{ongoing}.  A bug is
open iff it has been discovered but not yet resolved internally. Once
a bug has been resolved internally, its fix will be deployed in a
future patch to the production servers.  The patches for the
components of the email service are pre-scheduled.  Selected relations
of our running example are shown in \Cref{fig:exampleOngoingRelations}
and discussed below.

\approachOngoingRelations

Relation $\mathbf{B}$ illustrates bugs described by identifier \emph{BID},
the name of the affected component $C$, the valid time interval $VT$
during which the bug is open, and the reference time \emph{RT} when
the tuple belongs to the instantiated relations (cf.\ below and
\Cref{sec:operators} for the details).
For instance, tuple $b_1$ records deprioritized bug $500$ for the
Spam filter component that has been open from $\dts{1}{25}$ until $\NowS$.

Relation $\mathbf{P}$ illustrates patches described by patch number
\emph{PID}, component \emph{C} to which the patch applies, valid time
interval \emph{VT} during which the patch is live, and the reference
time \emph{RT}.  For instance, tuple $p_1$ states that patch 201 of
the Spam filter is live from $\dts{8}{15}$ until $\dts{8}{24}$
exclusively.

Relation $\mathbf{L}$ lists the technical leads.  A technical lead is
described by their name, component \emph{C} they are responsible for,
valid time interval \emph{VT} during which they are responsible for
the component, and the reference time \emph{RT}.  For instance, tuple
$l_2$ records that Bob is the technical lead for the Spam filter
component from $\dts{8}{18}$ until $\NowS$.

Relations $\mathbf{B}$, $\mathbf{P}$, and $\mathbf{L}$ are \emph{base
  ongoing relations}.  All tuples belong to the instantiated relations
at all reference times and have a trivial reference time, i.e.,
$RT = \rtattribute{(-\infty, \infty)}$.  The reference time is
restricted by predicates on ongoing attributes. We will discuss the
restriction of a tuple's reference time in the following.

To schedule bug fixes, reprioritize bugs, and assess unresolved bugs,
we run a query that joins bugs that affect the Spam filter with
upcoming patches and technical leads:
\begin{align*}
& \mathbf{V} \leftarrow \pi_{BID, \mathbf{B}.VT, PID,
    \text{Name}, \mathbf{B}.VT \gv{\cap} \mathbf{L}.VT}( \\
& \hspace*{1.5cm}
  \sigma_{C = '\text{Spam filter}'}(\mathbf{B}) \\
& \hspace*{1.8cm}
  \Join_{(\mathbf{B}.C = \mathbf{P}.C)
     \wedge (\mathbf{B}.VT\ \gv{\text{before}}\ \mathbf{P}.VT)} \mathbf{P} \\
& \hspace*{1.8cm} \Join_{(\mathbf{B}.C = \mathbf{L}.C)
  \wedge (\mathbf{B}.VT\ \gv{\text{overlaps}}\ \mathbf{L}.VT)} \mathbf{L})
\end{align*}

We illustrate the computation of the reference time $RT$ for
$b_1 \Join_{\theta} p_1$ with
$\theta = ((\mathbf{B}.C = \mathbf{P}.C) \wedge (\mathbf{B}.VT\
\gv{\text{before}}\ \mathbf{P}.VT))$.  Conceptually, all occurrences
of $\NowS$ in predicate $\theta(b_1, p_1)$ are replaced with each
possible reference time $\tcurr$ in turn and the predicate is
evaluated.  This yields the following results for the \textit{before}
predicate:
\begin{center} \footnotesize
  \begin{tabular}{c|ccc}
      $\tcurr$
      & $[\dts{1}{25}, \NowS)$
      & $[\dts{8}{15}, \dts{8}{24})$
      & $b_1.VT\ \gv{\text{before}}\ p_1.VT$
      \\
      \hline
      ... & ... & ... & ... \\
      $\dts{8}{14}$
      & $[\dts{1}{25}, \dts{8}{14})$
      & $[\dts{8}{15}, \dts{8}{24})$
      & \emph{true} \\
      $\dts{8}{15}$
      & $[\dts{1}{25}, \dts{8}{15})$
      & $[\dts{8}{15}, \dts{8}{24})$
      & \emph{true} \\
      $\dts{8}{16}$
      & $[\dts{1}{25}, \dts{8}{16})$
      & $[\dts{8}{15}, \dts{8}{24})$
      & \emph{false}\\
      ... & ... & ... & ...
  \end{tabular}
\end{center}
At all reference times when the join predicate evaluates to
\emph{true}, the result tuple belongs to the instantiated relations.
In our example these are all reference times from $\dts{1}{26}$ up to $\dts{8}{15}$ and
we get $RT = \rtattribute{[\dts{1}{26}, \dts{8}{16})}$.

\approachOverviewBeforeQuery

Query result $\mathbf{V}$ includes the tuples illustrated in
\Cref{fig:exampleMatViewBefore}.  Note that (1) all ongoing time
points are preserved in $\mathbf{V}$.  For instance, the value of the
$\mathbf{B}.VT$ attribute makes it possible to identify prioritized
and deprioritized bugs. (2) The intersection
$\mathbf{B}.VT\ \gv{\cap}\ \mathbf{L}.VT$ states when a technical lead
is responsible for a bug. Consider tuple $v_1$ with
$b_1.VT\ \gv{\cap}\ l_1.VT = [\dts{1}{25}, \upNow{\dts{8}{18}})$,
which is an ongoing time interval.  Tuple $v_1$ states that Ann is the
responsible technical lead for bug 500 from $\dts{1}{25}$ until
possibly earlier, but not later than $\dts{8}{17}$.
Clearly, fixed time points together with $\NowS$ are not sufficient to
represent such results.  (3) The reference time of a tuple is
restricted by predicates on ongoing attributes. For each operator, the
reference time of the result tuples is determined by the reference
times when the input tuples belong to the instantiated relations
\emph{and} the reference times when the predicate evaluates to
\emph{true}. The reference time of the input tuples is relevant since
it is the result of predicates in earlier operators that derive these
tuples.  For instance, the reference time of the result tuples of join
$\sigma_{C = '\text{Spam filter}'}(\mathbf{B}) \Join_{\gv{\theta}}
\mathbf{P}$ was restricted by join predicate $\theta$. These tuples
are then input tuples for the join with ongoing relation $\mathbf{L}$.

\tableOngoingTimePoints

\section{Related Work}
\label{sec:relWork}

The most commonly used ongoing time point is $\NowS$.  The
state-of-the-art approach to deal with ongoing time points is to
instantiate them, i.e., replace them with the reference time.
Commercial database systems use the compile time as the reference time
whereas research approaches usually use the evaluation time as the
reference time.  Below we discuss the implications of both choices for
storing ongoing time points, query processing, and the validity of
query results.

Existing database systems cannot store ongoing time points.  They
instantiate ongoing time points immediately at compile time when
statements are issued.  The SQL-92 standard~\cite{SQL92} includes the
reserved keywords CURRENT\_TIME, CURRENT\_DATE, and CURRENT\_TIMESTAMP
that denote the ongoing time point $\NowS$ for different time
granularities.  These constructs can be used in SQL statements, but
are instantiated immediately at compile time.

Various research approaches have progressed the basic solution offered
by commercial database systems.  The key idea is to store ongoing time
points and instantiate them when accessing the data during query
processing.  The advantage of instantiating ongoing time points is
that existing query processing techniques can be used since the
instantiation eliminates ongoing time points \cite{temporalDbBook,
  efficientJoin, spatiotemporal, multiDim, notionsUpward, tempMod,
  temporalAlignment}.  The disadvantage is that query results are only
valid at the chosen reference time and get outdated by time passing
by.  Below we discuss different aspects of the instantiation that have
been investigated \cite{now, torp2000effective, finger1996semantics,
  transactionTimestamping}.  Throughout, we use $\timeDom$ to denote
the domain of fixed time points.

Clifford et al.~\cite{now, nowGeneral} propose a solution that handles
ongoing time point $\NowS$ during query processing.  Their framework
instantiates $\NowS$ whenever it is accessed. Thus, queries are
evaluated on instantiated relations without ongoing time points.  This
yields result relations that are only valid at the time when $\NowS$
was accessed.

Anselma et al.~\cite{nowQuerying} propose an algebra for relations
with ongoing time points.  Their goal is an approach that copes with
four commonly used representations of $\NowS$: \emph{Min},
\emph{Max}~\cite{layeredTemporal}, \emph{Null}, and \emph{Empty
  Range}~\cite{pointApproach,pointApproachBiTemp}.  Their time domain
is $\timeDom \cup \{\NowS\}$. They introduce intersection and
difference functions that may keep ongoing time points uninstantiated.
For instance, ongoing time points are not instantiated when the
resulting time interval contains $\NowS$ as end point like in
$[\dts{10}{14}, \NowS) \cap [\dts{10}{17}, \NowS) = [\dts{10}{17},
\NowS)$.  Their approach must instantiate $\NowS$ for more complex end
points.  For instance, $[\dts{10}{17}, \dts{10}{22})\ \cap$
$[\dts{10}{17}, \NowS) = [\dts{10}{17}, \dts{10}{20})$ at reference
time $\dts{10}{20}$.  Anselma et al.\ \cite{comprehensiveNow} have
extended their approach to support indeterminacy for tuples with
$\NowS$.  They have not worked out how predicates on ongoing time
points are defined and evaluated.

Snodgrass \cite{snodgrass1987temporal} proposes \emph{Forever} instead
of the ongoing time point $\NowS$.  \emph{Forever} denotes the largest
time point in the time domain, which is a fixed time point. Existing
query evaluation approaches for relations without ongoing time points
can be used on relations that use \emph{Forever}.  However, replacing
$\NowS$ with \emph{Forever} leads to incorrect results.  For instance,
at reference time $\dts{5}{14}$ the query \emph{``Which bugs might be
  resolved before patch 201 goes live?''} is not answered correctly.
Evaluating the query on relations $\mathbf{B}$ and $\mathbf{P}$ of
\Cref{fig:exampleOngoingRelations} with \emph{Forever} as the end
point results in bug 500 not being part of the result relation, which
is not correct.

Torp et al.~\cite{modificationSemanaticsNow} propose a solution for
modifications of temporal databases.  They show that performing
temporal modifications on tuples that are instantiated when accessed
leads to incorrect modifications and thus, incorrect data in the
database.  The authors propose time domain
$\mathcal{T}_f = \timeDom \cup \{\min(a,\NowS) | a \in \timeDom\} \cup
\{\max(a,\NowS) | a \in \timeDom\}$ to handle such modifications.
Instead of $\NowS$, they use the minimum and maximum of a time point
and $\NowS$ to correctly modify the database.  Time domain $T_f$
supports intersection and difference functions that do not instantiate
ongoing time points. Torp et al. use these two functions to express
temporal modifications that remain valid as time passes by. Their
approach cannot evaluate predicates on uninstantiated time attributes.
Queries with such predicates resort to Clifford's approach.  Thus,
query results get invalidated by time passing by.

Moving objects \cite{guting2005moving} change their spatial position
as time advances.  Research approaches in this area deal with
different types of queries on moving objects: static queries
\cite{movingObjectsSnapshot, movingObjectsWorkloadSnapshot},
continuous queries \cite{movingObjectsContinous, movingObjectsKNN,
  movingObjectsSkyline,
  movingObjectsFramework}, and time-parametrized queries
\cite{tpspatialqueries}.  \emph{Static queries} instantiate the moving
objects at a given reference time and are evaluated at fixed spatial
positions.  These approaches are similar to the approach of Clifford
et al.~\cite{now}, which instantiates ongoing time points.
\emph{Continuous queries} compute results that remain valid for a
short time span, e.g., 10 seconds, before the query is
re-evaluated. The results are continuously returned to applications.
A query result contains pairs of moving object(s) and the reference times
when the pair belongs to the result.  Structurally, the query result
is similar to ongoing relations with a reference time attribute.
However, the result of a continuous query is only valid for a short
time span and gets invalidated by time passing by.
\emph{Time-parametrized queries} \cite{tpspatialqueries} incrementally
determine their results.  The result consists of three parts: the
objects that satisfy the spatial query, the reference time until when
the result is valid, and the objects that change the result. The
result is only valid from the time when the query was issued until the
returned reference time.

Now-relative and indeterminate time points have been proposed as
extensions of ongoing time point $\NowS$ \cite{now}.  A
now-relative time point, e.g., $\NowS + 5\ \text{days}$, shifts
$\NowS$ by 5 days into the future. An indeterminate time point
specifies a period during which an event will occur.  For instance,
the indeterminate time point $\dts{4}{17} \sim \dts{4}{20}$ as the end
point of a resolved bug states that the resolution occurred sometime
between $\dts{4}{17}$ and $\dts{4}{20}$.  These extensions are
orthogonal to our generalization of $\NowS$.

\section{Preliminaries}
\label{sec:background}

We assume a linearly ordered, discrete \emph{time domain} $\timeDom$
with $-\infty$ as the lower limit and $\infty$ as the upper limit.  A
\emph{time point} is an element of time domain $\timeDom$.  A
\emph{time interval} $[t_s, t_e)$ consists of an inclusive start point
$t_s$ and an exclusive end point $t_e$.  \emph{Fixed} data types
consist of values that do not change as time passes by.  Examples are
integers, strings, booleans, and time points of $\timeDom$.
\emph{Ongoing} data types include values that change as time passes
by.  Ongoing values can be \emph{instantiated} to fixed values.  We
consider the following ongoing data types: ongoing time points,
ongoing booleans, and composite structures (intervals, tuples,
relations) that include ongoing time points.  The bind operator
$\bind{x}{\tcurr}$ performs the instantiation of $x$ at reference time
$\tcurr \in \timeDom$.  If $x$ is composite each component is
instantiated.
We use the $\fixed{}$-superscript for operations on fixed data types.
For instance, $\minF$ is the standard minimum function over fixed
arguments, i.e., $\minF(j,k) = j$ if $j < k$ and $\minF(j,k) = k$
otherwise.

$R = (\mathbf{A})$ denotes the schema of a fixed relation $\mathbf{R}$
with fixed attributes $\mathbf{A} = A_1, \dots, A_n$.  A tuple $r$
with schema $R$ is a finite list that contains for every $A_i$ a value
from the domain of $A_i$. A relation $\mathbf{R}$ over schema $R$ is a
finite set of tuples over $R$.  $r.A_i$ denotes the value of attribute
$A_i$ in tuple $r$.  $\theta(r)$ denotes the application of predicate
$\theta$ to tuple $r$.  An \emph{ongoing relation} is a relation with
fixed and ongoing attributes $\mathbf{A}$ and a reference time
attribute $RT$ (cf. \Cref{def:1}).  The value of $RT$ is a set of
fixed time intervals.

Valid time \cite{temporalGlossary}, transaction time
\cite{temporalGlossary}, and reference time are separate concepts.
Consider a tuple $b$ that refers to bug $500$ with valid time
$VT = [\dts{1}{25}, \NowS)$, transaction time
$TT = [\dts{1}{26}, \NowS)$, and reference time
$RT = \rtattribute{[\dts{3}{15}, \infty)}$.  The valid time states
when a tuple is valid in the real world: bug 500 is open from
$\dts{1}{25}$ until $\NowS$.  The valid time is set by the user.  The
transaction time states when a tuple was modified in the relation:
tuple $b$ was inserted in $\dts{1}{26}$ and not modified since.  The
transaction time is restricted by the database system through database
modifications, i.e., insert, update, and delete statements.  The
reference time states when a tuple belongs to the instantiated
relations: tuple $b$ belongs to the instantiated relations from
$\dts{3}{15}$ on.  The reference time is set by the database system
and restricted by the predicates on ongoing attributes in queries.

%

\section{Ongoing Time Data Types}
\label{sec:ongoingDataTypes}

This section defines the ongoing time domain $\ongoingDom$, ongoing
time points, and ongoing time intervals.  In contrast to previously
proposed ongoing time domains, $\ongoingDom$ is closed for minimum and
maximum functions \crc{(cf. Proof of \Cref{th:coreOpsEquivalences}).}

\tableOngoingIntervals

\subsection{Ongoing Time Points}
\label{sec:ongoing-time-points}

\begin{definition}[Ongoing Time Domain $\ongoingDom$]
	\label{def:timeDomain}
		Let $\timeDom$ be the time domain of fixed time points.
		Ongoing time domain $\ongoingDom$ consists of all possible ongoing
		time points $\minPoint{a}{b}$:
		\begin{equation*}
		\ongoingDom = \{\minPoint{a}{b}\ |\ \exists a, b \in \timeDom (a \leq b)\}
		\end{equation*}
\end{definition}

The intuitive meaning of the ongoing time point
$\minPoint{a}{b}$ is \emph{not earlier than $a$, but not later than
  $b$}. For instance,
$\minPoint{\dts{10}{17}}{\dts{10}{19}}$ means \emph{not
  earlier than $\dts{10}{17}$, but not later than
  $\dts{10}{19}$}.

\begin{definition}[Ongoing Time Point]
  \label{def:ongoingTimePoint}
  Let $\tcurr \in \timeDom$ be a
  reference time and $a,b \in \timeDom$ with $a \leq b$. The ongoing
  time point $\minPoint{a}{b}$ is defined as
  \begin{align*}
    \bind{\minPoint{a}{b}}{\tcurr} = \begin{dcases*}
      a & $\tcurr \leq a$\\
      \tcurr & $a < \tcurr < b$ \\
      b & otherwise
    \end{dcases*}
  \end{align*}
\end{definition}
For instance, ongoing time point
$\minPoint{\dts{10}{17}}{\dts{10}{19}}$ instantiates to time point
$\dts{10}{17}$ up to reference time $\dts{10}{17}$. Between reference
times $\dts{10}{17}$ and $\dts{10}{19}$ the ongoing time point
instantiates to the reference time. Afterwards, it instantiates to
time point $\dts{10}{19}$.

A \emph{fixed} time point $a$, current time point $\NowS$, a
\emph{growing} time point $\Now{a}$, and a \emph{limited} time point
$\upNow{b}$ can all be expressed as ongoing time points of the form
$\minPoint{a}{b}$.  This is illustrated in \Cref{fig:timeDomain}.
For instance, fixed time point $a = \minPoint{a}{a}$ is an ongoing
time point that instantiates to time point $a$ at all reference
times; time point $\NowS = \minPoint{-\infty}{\infty}$ is an ongoing
time point that instantiates to the reference time at all reference
times.

\Cref{tab:overviewTimeDomains} summarizes the properties of time
domains $\timeDom$, $\timeDom_{\NowS} = \timeDom \cup \{\NowS\}$
\cite{now},
$\mathcal{T}_f = \timeDom \cup \{\min(a,\NowS) |\ a \in \timeDom\}
\cup \{\max(a,\NowS) |\ a \in \timeDom\}$
\cite{modificationSemanaticsNow}, and $\ongoingDom$.  For each time
domain we show if it includes fixed or ongoing time points and if it
is closed for $\minV$ and $\maxV$.

\begin{table}[!htb] \centering
	\caption{Properties of time domains.}
	\label{tab:overviewTimeDomains}
	\begin{tabular}{c|ccc}
		\hline
		Time Domain & Fixed & Ongoing & Closed \\ \hline
		$\timeDom$ & yes & no & yes \\
		$\timeDom_{\NowS}$ & yes & yes & no \\
		$\timeDom_f$ & yes & yes & no \\
		$\mathbf{\ongoingDom}$ & \textbf{yes} & \textbf{yes} & \textbf{yes} \\
		\hline
	\end{tabular}
\end{table}

\subsection{Ongoing Time Intervals}

An ongoing time interval $[t_s, t_e)$ is a closed-open time interval
with domain $\ongoingDom \times \ongoingDom$.  As an example, time
interval $[\dts{10}{17},\NowS)$ is an ongoing time interval. An
  ongoing time interval can be instantiated to a fixed time interval
  by instantiating start and end points:
\[\forall \tcurr \in \timeDom (\bind{[t_s, t_e)}{\tcurr} = [\bind{t_s}{\tcurr}, \bind{t_e}{\tcurr}))\]

The ongoing time interval $[\minPoint{a}{b}, \minPoint{c}{d})$
generalizes \emph{fixed} time intervals, \emph{expanding} time
intervals, and \emph{shrinking} time intervals. Their semantics are
illustrated in \Cref{fig:ongoingTimeIntervals}.  For instance, an
expanding time interval instantiates to time intervals whose duration
increases with increasing reference time. The duration can increase
for all reference times or up to a certain reference time. An example
for the first case is ongoing time interval
$[\dts{10}{17}, \NowS)$ with $d = \infty$. An example for the latter
case is ongoing time interval
$[\dts{10}{17}, \minPoint{\dts{10}{19}}{\dts{10}{21}})$ with
$d = \dts{10}{21}$.  It instantiates to time intervals with increasing
duration up to reference time $\dts{10}{21}$. From reference
  time $\dts{10}{21}$ on, it instantiates to time interval
  $[\dts{10}{17}, \dts{10}{21})$.

An ongoing time interval can be partially empty.  A partially empty
time interval instantiates to empty time intervals at some reference
times and to non-empty time intervals at others.
This is illustrated in \Cref{fig:ongoingTimeIntervals}.
For instance,
ongoing time interval $[\dts{10}{17}, \NowS)$ instantiates to empty time
intervals up to reference time $\dts{10}{17}$.  At these reference times,
end point $\NowS$ instantiates to time points that are less than or
equal to start point $\dts{10}{17}$ and the interval is empty.  Afterwards,
$\NowS$ instantiates to time points greater than $\dts{10}{17}$ and
$[\dts{10}{17}, \NowS)$ instantiates to non-empty time intervals.

\section{Operations on Ongoing Data Types}
\label{sec:gvOperations}

This section defines operations, i.e., functions, predicates, and
  logical connectives, on ongoing time data types whose results
remain valid as time passes by.  At each reference time, their results
are equal to the results obtained by the corresponding operation on
fixed data types.
We provide and prove equivalences
for our six core operations $<, \minV, \maxV, \land, \lor, \neg$
and show how we use these core operations in equivalences for
additional operations on ongoing data types.

Since ongoing time points and time intervals instantiate to
different values depending on the reference time the truth value of
predicates depends on the reference time.  To represent their result,
we use \emph{ongoing booleans} whose boolean value depends on the
reference time.

\begin{definition}[Ongoing Boolean]
	\label{def:ongoingBoolean}
	Let $\tcurr \in \timeDom$ be a reference time. Let
    $S_t \subseteq \timeDom$ and $S_f \subseteq \timeDom$ be disjoint
    subsets of all possible reference times with
  $S_t \cup S_f = \timeDom$.  The ongoing boolean
  $\ongoingBoolVar{S_t}{S_f}$ is defined as
	\begin{align*}
	\bind{\ongoingBoolVar{S_t}{S_f}}{\tcurr} =
	\begin{dcases}
	\text{true} & \tcurr \in S_t \\
	\text{false} & \tcurr \in S_f
	\end{dcases}
	\end{align*}
\end{definition}
An ongoing boolean $\ongoingBoolVar{S_t}{S_f}$ is \emph{true} at
the reference times in $S_t$ and \emph{false} at the reference times
in $S_f$.  For instance, ongoing boolean
$\ongoingBool{[\dts{10}{18}, \infty)}{(-\infty, \dts{10}{18})}$ is
\emph{true} at reference time $\dts{10}{18}$ (as well as at all
later reference times), and it is \emph{false} at the reference
times earlier than $\dts{10}{18}$.
Ongoing booleans generalize booleans.  Boolean \emph{true} is
equivalent to ongoing boolean
$\ongoingBool{(-\infty,\infty)}{\emptyset}$, which is \emph{true} at
all reference times.  Boolean \emph{false} is equivalent to ongoing
boolean $\ongoingBool{\emptyset}{(-\infty,\infty)}$.  The
generalization makes it possible to combine predicates that evaluate
to booleans with predicates that evaluate to ongoing booleans in
logical expressions.

\begin{definition}[Core Operations]
	\label{fig:coreOperations}
	Let $t_1, t_2, t \in \ongoingDom$ be ongoing time points. Let
  $\ongoingBoolLetter{1}, \ongoingBoolLetter{2}, \ongoingBoolLetter{}
  \in \boolDom$ be ongoing booleans. The core operations on ongoing
  data types are defined as follows:\smallskip
	\begin{center}
		\footnotesize
		\renewcommand{\arraystretch}{1.6}
		\setlength\arraycolsep{1pt}
		\begin{centermath}
			\begin{array}{cl}
				\hline
				\textbf{Operation} & \hspace{1pt}\textbf{Definition} \\ \hline
				\gv{<}
				& t_1 \gv{<} t_2 = \ongoingBoolLetter{} \text{ iff } \forall \tcurr \in \mathcal{T} (
				\begin{aligned}[t]
					& \bind{\ongoingBoolLetter{}}{\tcurr} \Leftrightarrow \bind{t_1}{\tcurr} \fixed{<} \bind{t_2}{\tcurr})
				\end{aligned}
				\\ \hline
				\minV
				& \minV(t_1, t_2) = t \text{ iff } \forall \tcurr \in \mathcal{T} (\bind{t}{\tcurr} = \minF(\bind{t_1}{\tcurr}, \bind{t_2}{\tcurr}))
				\\
				\maxV
				& \maxV(t_1, t_2) = t \text{ iff } \forall \tcurr \in \mathcal{T} (\bind{t}{\tcurr} = \maxF(\bind{t_1}{\tcurr}, \bind{t_2}{\tcurr}))
				\\ \hline
				\gv{\wedge}
				& \ongoingBoolLetter{1} \gv{\wedge} \ongoingBoolLetter{2} = \ongoingBoolLetter{} \text{ iff } \forall \tcurr \in \mathcal{T} (\bind{\ongoingBoolLetter{}}{\tcurr} \Leftrightarrow \bind{\ongoingBoolLetter{1}}{\tcurr} \fixed{\wedge} \bind{\ongoingBoolLetter{2}}{\tcurr})
				\\
				\gv{\vee}
				& \ongoingBoolLetter{1} \gv{\vee} \ongoingBoolLetter{2} = \ongoingBoolLetter{} \text{ iff } \forall \tcurr \in \mathcal{T} (\bind{\ongoingBoolLetter{}}{\tcurr} \Leftrightarrow \bind{\ongoingBoolLetter{1}}{\tcurr} \fixed{\vee} \bind{\ongoingBoolLetter{2}}{\tcurr})
				\\
				\gv{\neg}
				& \gv{\neg}\ongoingBoolLetter{1} = \ongoingBoolLetter{} \text{ iff } \forall \tcurr \in \mathcal{T} (\bind{\ongoingBoolLetter{}}{\tcurr} \Leftrightarrow \fixed{\neg} \bind{\ongoingBoolLetter{1}}{\tcurr})
				\\ \hline
			\end{array}
		\end{centermath}
	\vspace*{5pt}
	\end{center}
\end{definition}
An operation on ongoing data types evaluates to a result that, at each
reference time, is equal to the result obtained by the corresponding
operation on fixed data types.  This yields results that remain valid
as time passes by.

All other predicates and functions on ongoing data types are defined
analogously.

\begin{example}
	Consider $\minV$ for ongoing time points and the corresponding
  function $\minF$ for fixed time points. The result of
  $\minV(\dts{10}{17}, \NowS)$ is ongoing time point
  $t = \upNow{\dts{10}{17}}$ (cf. \Cref{th:coreOpsEquivalences}). At
  each reference time, it is equal to the time point obtained from
  evaluating $\minF$ on the instantiated input arguments, i.e.,
  $\upNow{\dts{10}{17}}$ is equal to
  $\minF(\bind{\dts{10}{17}}{\tcurr}, \bind{\NowS}{\tcurr})$ at every
  reference time $\tcurr$. \Cref{fig:generalValidityExample}
  illustrates the equality for reference times $\dts{10}{15}$ and
  $\dts{10}{19}$.

	\begin{figure}[!htb] \centering
		\begin{tikzpicture}[shorten >=1pt,auto, node distance = 3cm,
		thick, scale=1, every node/.style={scale=0.8}]
		\tikzstyle{transition2}=[rectangle, thick, draw=black!20, minimum size=4mm]
		\def\xa{2.1}
		\def\xb{4.1}
		\def\xc{6.1}

		\draw (\xb,-0.2) node[above, anchor=south, align=center]
		{$\bind{\upNow{\dts{10}{17}}}{\tcurr}$};
		\draw[->] (0,-0.2) -- (8.2,-0.2) node[below] {$\tcurr$};
		\draw (\xa,-0.3) -- (\xa, -0.06);
		\draw (\xc,-0.3) -- (\xc, -0.06);

		\draw (\xa,-0.25) node[below] {$\dts{10}{15}$};
		\draw (\xc,-0.25) node[below] {$\dts{10}{19}$};
		\draw[->, darkred] (\xa, -0.6) -- (\xa, -1.);
		\draw[->, darkred] (\xc, -0.6) -- (\xc, -1.);
		\draw (\xa,-1) node[below, anchor=north, transition2, align=left]{
			$\mathcoloring{lightBlue}{\bind{\upNow{\dts{10}{17}}}{\dts{10}{15}}}$
			$= \mathbf{\dts{10}{15}}$
		};
		\draw (\xc,-1) node[below, anchor=north, transition2, align=left]{
			$\mathcoloring{lightBlue}{\bind{\upNow{\dts{10}{17}}}{\dts{10}{19}}} = \mathbf{\dts{10}{17}}$
		};
		\draw (\xa, -1.55) node[below] {$\mathbf{=}$};
		\draw (\xc, -1.55) node[below] {$\mathbf{=}$};
		\draw (\xa,-1.9) node[below, anchor=north, transition2, align=left]{
			$\mathcoloring{lightBlue}{\minF(\dts{10}{17},\dts{10}{15})}$
			$= \mathbf{\dts{10}{15}}$
		};
		\draw(\xc,-1.9) node[below, anchor=north, transition2, align=left]{
			$\mathcoloring{lightBlue}{\minF(\dts{10}{17},\dts{10}{19})}$
			$= \mathbf{\dts{10}{17}}$
		};
		\draw[->, darkred] (\xa, -2.85) -- (\xa, -2.45);
		\draw[->, darkred] (\xc, -2.85) -- (\xc, -2.45);
		\draw (\xa,-3.2) node[above] {$\dts{10}{15}$};
		\draw (\xc,-3.2) node[above] {$\dts{10}{19}$};
		\draw (\xa,-3.2) -- (\xa, -3.45);
		\draw (\xc,-3.2) -- (\xc, -3.45);
		\draw[->] (0,-3.3) -- (8.2,-3.3) node[above] {$\tcurr$} ;
		\draw (\xb, -3.4) node[below, anchor=north, align=left] {
			$\mathbf{\minF(\bind{\dts{10}{17}}{\tcurr}, \bind{\NowS}{\tcurr})}$
		};
		\end{tikzpicture}
		\caption{The result of $\minV$ remains valid.}
		\label{fig:generalValidityExample}
	\end{figure}
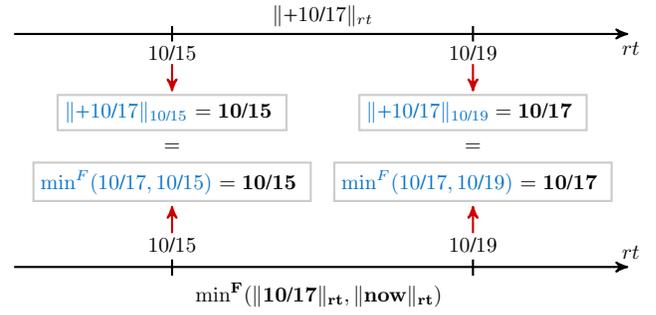
\end{example}

\begin{lemma}
	\label{th:coreOpsEquivalences}
	Let $\minPoint{a}{b}, \minPoint{c}{d} \in \ongoingDom$ be ongoing
  time points.  Let
  $\ongoingBoolVar{S_t}{S_f},
  \ongoingBoolVar{\tilde{S}_t}{\tilde{S}_f} \in \boolDom$ be ongoing
  booleans.  The results of the operations on ongoing data types
  given in \Cref{fig:coreOperations} are equivalent to the
    following ongoing values:
	\begin{center}
	\vspace*{5pt}
	\footnotesize
	\renewcommand{\arraystretch}{1.6}
	\begin{centermath}
		\begin{array}{cl}
			\hline
			\textbf{Operation} & \textbf{Equivalence} \\ \hline
			\gv{<}
			&
				\begin{aligned}[t]
					&\minPoint{a}{b} \gv{<} \minPoint{c}{d}\ \equiv \\
					&\hspace*{-.7cm}
					\begin{dcases}
						\ongoingBool{(-\infty, \infty)}{\emptyset} & a \leq b < c \leq d\ \ (1)\\
						\ongoingBool{(-\infty, c)}{[c, \infty)} & a < c \leq d \leq b\ \ (2)\\
						\ongoingBool{[b + 1, \infty)}{(-\infty, b+1)} & c \leq a \leq b  < d\ \ (3)\\
						\mathbf{b}[\{(-\infty, c), [b+1, \infty)\},\{[c, b+1)\} & a < c \leq b < d\ \ (4)\\
						\ongoingBool{\emptyset}{(-\infty, \infty)} & \text{otherwise}\ \hspace*{22pt} (5)
					\end{dcases}
				\end{aligned}
			\\ \hline
			\minV
			& \minV(\minPoint{a}{b}, \minPoint{c}{d}) \equiv \minPoint{\minF(a,c)}{\minF(b,d)}
			\\
			\maxV
			&\maxV(\minPoint{a}{b}, \minPoint{c}{d}) \equiv \minPoint{\maxF(a,c)}{\maxF(b,d)}
			\\ \hline
			\gv{\wedge}
			&\ongoingBoolVar{S_t}{S_f} \gv{\wedge} \ongoingBoolVar{\tilde{S}_t}{\tilde{S}_f} \equiv \ongoingBoolVar{S_t \fixed{\cap} \tilde{S}_t}{S_f \fixed{\cup} \tilde{S}_f}
			\\
			\gv{\vee}
			&\ongoingBoolVar{S_t}{S_f} \gv{\vee} \ongoingBoolVar{\tilde{S}_t}{\tilde{S}_f} \equiv \ongoingBoolVar{S_t \fixed{\cup} \tilde{S}_t}{S_f \fixed{\cap} \tilde{S}_f}
			\\
			\gv{\neg}
			&\gv{\neg} \ongoingBoolVar{S_t}{S_f} \equiv \ongoingBoolVar{S_f}{S_t}
			\\ \hline
		\end{array}
	\end{centermath}
	\end{center}
\end{lemma}


\crc{
\begin{IEEEproof}
	We prove the equivalences in the order provided in
  \Cref{th:coreOpsEquivalences}.

		The equivalence for $\minPoint{a}{b} \gv{<} \minPoint{c}{d}$
		is proven by showing for each ordering of $a$, $b$, $c$, and
		$d$ that the definition of $\gv{<}$ holds (cf.  \Cref{fig:coreOperations}).
		We show for the ordering $a < c = d < b$ that ongoing boolean
		$\ongoingBool{(-\infty, c)}{[c, \infty)}$ (case 2 of the equivalence)
		fulfills the definition, i.e.,
		$\forall \tcurr \in \timeDom (\bind{\ongoingBool{(-\infty, c)}{[c, \infty)}}{\tcurr} \Leftrightarrow
		\bind{\minPoint{a}{b}}{\tcurr} \fixed{<} \bind{\minPoint{c}{d}}{\tcurr})$.
		The following table shows that the definition is fulfilled at every reference
		time: $\fixed{<}$ evaluates to the same boolean as ongoing boolean $\ongoingBoolLetter{}  = \ongoingBool{(-\infty, c)}{[c, \infty)}$ instantiates to.

		{
		\footnotesize
		\renewcommand{\arraystretch}{1.2}
		\begin{centermath}
			\begin{array}{ccccc}
				\tcurr & \bind{\minPoint{a}{b}}{\tcurr} & \bind{\minPoint{c}{d}}{\tcurr} & \fixed{<} & \bind{\ongoingBoolLetter{}}{\tcurr} \\ \hline
				\mathbf{\tcurr} \leq a < c = d < b & a & c & \text{true} & \text{true} \\
				a < \mathbf{\tcurr} < c = d < b & \tcurr & c & \text{true} & \text{true} \\
				a < \mathbf{\tcurr} = c = d < b & \tcurr & c & \text{false} & \text{false} \\
				a < c = d < \mathbf{\tcurr} < b & \tcurr & c & \text{false} & \text{false} \\
				a < c = d < b \leq \mathbf{\tcurr} & b & c & \text{false} & \text{false} \\
				\hline
			\end{array}
		\end{centermath}
		}
		The equivalence is proven for the other orderings analogously.

	We prove $\minV(\minPoint{a}{b}, \minPoint{c}{d}) \equiv \minPoint{\minF(a,c)}{\minF(b,d)}$ by showing that
	\begin{enumerate*}[label=(\arabic*)]
		\item $\minPoint{\minF(a,c)}{\minF(b,d)}$ is an ongoing time point of $\ongoingDom$, and
		\item the definition of $\minV$ (cf. \Cref{fig:coreOperations}) holds for $t = \minPoint{\minF(a,c)}{\minF(b,d)}$.
	\end{enumerate*}
	Let $\minPoint{a}{b}, \minPoint{c}{d} \in \ongoingDom$ be two ongoing time points with $a \leq b$ and $c \leq d$.
	First, for fixed values, $\minF(a,c) \leq a$ and $\minF(a,c) \leq c$ hold. To prove
	$\minPoint{\minF(a,c)}{\minF(b,d)} \in \ongoingDom$ we must
	show that $\minF(a,c) \leq \minF(b,d)$.
	\begin{align*}
	& \text{Case 1: }
	\begin{aligned}[t]
	& \minF(b,d) = b \\
	& \minF(a,c) \leq a \leq b = \minF(b,d)
	\end{aligned}\\
	& \text{Case 2: }
	\begin{aligned}[t]
	& \minF(b,d) = d \\
	& \minF(a,c) \leq c \leq d = \minF(b,d)
	\end{aligned}
	\end{align*}
	Second, we show that the definition of $\minV$ holds for $t = \minPoint{\minF(a,c)}{\minF(b,d)}$. Let $\tcurr \in \mathcal{T}$ be a reference time. From \Cref{def:ongoingTimePoint} it follows that the instantiation
	$\bind{\minPoint{a}{b}}{\tcurr}$ is
	equivalent to  $\minF(b, \maxF(a, \tcurr))$.
	\begin{align*}
	& \bind{t}{\tcurr} =
	\minF(\bind{\minPoint{a}{b}}{\tcurr}, \bind{\minPoint{c}{d}}{\tcurr})
	\\
	&\Leftrightarrow
	\hspace{4pt}
	\begin{aligned}[t]
	&\bind{\minPoint{\minF(a,c)}{\minF(b,d)}}{\tcurr} \\
	&= \minF(\minF(b, \maxF(a, \tcurr)),\minF(d, \maxF(c, \tcurr)))
	\end{aligned}
	\\
	& \Leftrightarrow\footnotemark
	\hspace{2pt}
	\begin{aligned}[t]
	&\minF(\minF(b,d), \maxF(\minF(a,c), \tcurr)) \\
	&= \minF(\minF(b, d), \minF(\maxF(a, \tcurr), \maxF(c, \tcurr)))
	\end{aligned}
	\\
	& \Leftrightarrow\footnotemark
	\hspace{2pt}
	\begin{aligned}[t]
	&\minF(\minF(b,d), \maxF(\minF(a,c), \tcurr)) \\
	&= \minF(\minF(b, d), \maxF(\minF(a, c), \tcurr))
	\end{aligned}
	\end{align*}
	Thus, $\minV(\minPoint{a}{b}, \minPoint{c}{d}) \equiv \minPoint{\minF(a,c)}{\minF(b,d)}$ holds.
	The equivalence of $\maxV$ is proven analogously.

	\footnotetext[1]{$\minF$ is associative, i.e., \\\hspace*{9pt} $\minF(\minF(w,x), \minF(y, z)) = \minF(\minF(w,y), \minF(x,z))$}
\footnotetext[2]{$\minF$ and $\maxF$ are distributive over each other,
	i.e.,\\
	$\hspace*{12pt}\minF(\maxF(x, z), \maxF(y, z)) = \maxF(\minF(x,y),
	z)$}

	The logical conjunction of two ongoing booleans is ongoing boolean
	$\ongoingBoolVar{S_t \fixed{\cap} \tilde{S}_t}{S_f \fixed{\cup}
		\tilde{S}_f}$. It instantiates to \emph{true} at the reference
	times when both input ongoing booleans instantiate to \emph{true}, i.e.,
	$S_t \fixed{\cap} \tilde{S}_t$; it instantiates to \emph{false} when
	at least one of the input ongoing booleans instantiate to \emph{false}, i.e.,
	at the union $S_f \fixed{\cup} \tilde{S}_f$.
	The disjunction of two ongoing booleans is ongoing boolean
	$\ongoingBoolVar{S_t \fixed{\cup} \tilde{S}_t}{S_f \fixed{\cap}
		\tilde{S}_f}$. It instantiates to \emph{true} at the reference
	times when at least one of the input ongoing booleans instantiates
	to \emph{true}.
	The negation of an ongoing boolean
	$\ongoingBoolVar{S_t}{S_f}$ is $\ongoingBoolVar{S_f}{S_t}$. This means
	that at the reference times when the input ongoing boolean
	instantiates to \emph{true}, the resulting ongoing boolean
	instantiates to \emph{false}.
\end{IEEEproof}
}

We use our core operations to provide equivalences for
predicates and functions on ongoing time points and time intervals.
\Cref{fig:remainingOps} illustrates the equivalences for selected
predicates and functions.  For instance, the intersection
$[t_s, t_e) \gv{\cap} [\tilde{t}_s, \tilde{t}_e)$ on ongoing time
intervals is equivalent to the ongoing time interval
$[\maxV(t_s,\tilde{t}_s), \minV(t_e,\tilde{t}_e))$.

\nonCoreOps

For predicates on ongoing time intervals we must explicitly consider
the non-emptiness of the ongoing time intervals.  For instance, the
$\text{overlaps}$ predicate is equivalent to the ongoing boolean that
results from the usual overlaps check
$t_s < \tilde{t}_e\ \land\ \tilde{t}_s < t_e$ and an explicit
non-empty check
$t_s \gv{<} t_e \gv{\wedge} \tilde{t}_s \gv{<} \tilde{t}_e$.
The explicit non-empty check is necessary because ongoing time intervals
can be partially empty.  It is not sufficient to check if the ongoing
input time intervals are not empty at all reference times; we must
check non-emptiness at each reference time.

\begin{example}
	Consider the $\gv{\text{overlaps}}$ predicate.  At all reference
	times when one of the input time intervals instantiates to an
	empty time interval, the non-empty check ensures that the predicate
	evaluates to \emph{false}.  At all other reference times, the
	overlaps check determines the result.  At reference time
	$\dts{10}{16}$, ongoing time interval
	$[\dts{10}{17}, \NowS)$ instantiates to an empty time interval
	and thus, predicate
	$[\dts{10}{17}, \NowS)\ \gv{\text{overlaps}}\
	[\dts{10}{14}, \dts{10}{20})$ evaluates to \emph{false}. At reference
	time $\dts{10}{18}$, both ongoing input time intervals
	instantiate to non-empty time intervals and the overlaps check
	evaluates to \emph{true}.  Thus, predicate
	$[\dts{10}{17}, \NowS)\ \gv{\text{overlaps}}\
	[\dts{10}{14}, \dts{10}{20})$ evaluates to ongoing boolean
	$\ongoingBool{[\dts{10}{18}, \infty)}{(-\infty,
		\dts{10}{18})}$.
\end{example}

\section{Relational Algebra}
\label{sec:operators}

The first subsection introduces \emph{ongoing relations} to represent
query results that remain valid at varying times. Ongoing relations include tuples
that belong to instantiated relations at some reference times only. An
ongoing relation models this by associating each tuple with a
reference time attribute. The value of the reference time attribute is
restricted by the predicates on ongoing attributes. The second
subsection defines the operators of the relational algebra as
operators on ongoing relations.

\subsection{Ongoing Relations}
\label{sec:ongoing-relations}

\begin{definition}[Schema of an Ongoing Relation]
	\label{def:1}
  Let $\mathbf{A}$ be a
  list of fixed and ongoing attributes $A_1, \dots,$ $A_n$ and
  $RT$ be the reference time attribute.  Then,
  \[R = (\mathbf{A}, RT)\] is the schema of an ongoing relation.
\end{definition}

A tuple belongs to the instantiated relations at the reference times
that are contained in the value of the tuple's reference time
attribute $RT$. In a base tuple, the value of $RT$ is set to trivial
reference times, i.e., $RT = \rtattribute{(-\infty, \infty)}$, by the
database system. The reference time of tuples is then restricted by
predicates on ongoing attributes.

The bind operator
$\bind{\mathbf{R}}{\tcurr}$ instantiates an ongoing relation
$\mathbf{R}$ at reference time $\tcurr \in \timeDom$ by instantiating
the ongoing attributes of each tuple at reference time $\tcurr$.  It
omits tuples whose reference time $RT$ does not contain $\tcurr$:
\[
  \bind{\mathbf{R}}{\tcurr} =
    \{x | \exists r \in \mathbf{R}( x.\mathbf{A}
  =
  \bind{r.\mathbf{A}}{\tcurr} \wedge \tcurr \in r.RT)\}
\]

\subsection{Operators on Ongoing Relations}
\label{sec:oper-ongo-relat}

The definition of the relational algebra operators on ongoing
relations follows the approach in \Cref{fig:coreOperations}.  For
instance, selection $\gv{\sigma}_{\gv{\theta}}(\mathbf{R})$ for
ongoing relations is defined as
\[ \gv{\sigma}_{\gv{\theta}}(\mathbf{R}) = \mathbf{V} \text{ iff }
  \forall \tcurr \in \timeDom (\bind{\mathbf{V}}{\tcurr} \equiv
  \fixed{\sigma}_{\fixed{\theta}}(\bind{\mathbf{R}}{\tcurr}))
\]

Derived relational algebra operators are defined as usual.  As an example,
$\mathbf{R} \gv{\Join}_{\gv{\theta}} \mathbf{S} =
\gv{\sigma}_{\gv{\theta}}(\mathbf{R} \gv{\times} \mathbf{S})$.

\begin{lemma}
	\label{th:raOperators}
	Let $\mathbf{R}, \mathbf{S}$ be two ongoing relations with attributes $\mathbf{A}$ and $\mathbf{C}$,
	  respectively. Let $\mathbf{B} \subseteq \mathbf{A}$ be a subset of
	  the attributes of $\mathbf{R}$ and let predicate
	  $\gv{\theta}$ be composed of operations whose results remain valid as time passes by (cf. \Cref{sec:gvOperations}).
	The results of the relational algebra operators on ongoing relations are equivalent to the following ongoing relations:

	 \vspace*{-5pt}
	{
		\centering
		\footnotesize
		\renewcommand{\arraystretch}{1.6}
		\setlength\arraycolsep{2pt}
		$$
		\begin{array}{lr@{\ \equiv \ }l}
		 \hline \textbf{Operator} & \multicolumn{2}{l}{\textbf{Equivalence}} \\ \hline
		\text{Projection} & \gv{\pi}_{\mathbf{B}}(\mathbf{R}) & \{x | \exists r \in \mathbf{R}(x.\mathbf{B} = r.\mathbf{B} \wedge x.RT = r.RT) \} \\
		\text{Selection} & \gv{\sigma}_{\gv{\theta}}(\mathbf{R}) &
		\begin{aligned}[t]
				&\{x | \exists r \in \mathbf{R}(x.\mathbf{A} = r.\mathbf{A} \wedge x.RT = (r.RT \gv{\wedge} \gv{\theta}(r)) \\ &
				\hspace*{12pt}\wedge x.RT
		\neq \emptyset)\}
		\end{aligned}
  \\
		\text{Cart. prod.} & \mathbf{R} \gv{\times} \mathbf{S} & \begin{aligned}[t]
		&\{x |
		\exists r \in \mathbf{R}, s \in \mathbf{S}(x.\mathbf{A} = r.\mathbf{A} \wedge x.\mathbf{C} = s.\mathbf{C} \\
		&\hspace*{12pt}\wedge x.RT = (r.RT \gv{\wedge} s.RT) \wedge x.RT \neq \emptyset)
		\}
		\end{aligned} \\
		\text{Union} & \mathbf{R} \gv{\cup} \mathbf{S} & \{x | x \in \mathbf{R} \vee x \in \mathbf{S}\} \\
		\text{Difference} & \mathbf{R} \gv{-} \mathbf{S} &
		\begin{aligned}[t]
		&\{x |
		\exists r \in \mathbf{R}( x.\mathbf{A} = r.\mathbf{A} \wedge x.RT \neq \emptyset \\
		&\hspace*{12pt}\wedge x.RT = \{\tcurr \in r.RT | \nexists s \in \mathbf{S} (\\
		&\hspace*{36pt}	\bind{r.\mathbf{A}}{\tcurr} \fixed{=} \bind{s.\mathbf{A}}{\tcurr} \wedge \tcurr \in s.RT)\})
		\}
		\end{aligned}\\ \hline
		\end{array}
		$$
		}
\end{lemma}

\crc{
  \begin{IEEEproof}
	We prove the equivalence for selection $\gv{\sigma}_{\gv{\theta}}(\mathbf{R})$.
	For the other operators, similar transformations from the reference time of result tuples to
	instantiated relations hold.

	Let $\mathbf{R}$ be an ongoing relation and $\gv{\theta}$ be a
	predicate with operations whose results remain valid as time passes by.  Let $\fixed{\sigma}$ be
	the selection for fixed relations and predicate $\fixed{\theta}$ be
	the predicate we get by replacing operations in
	$\theta$ with the corresponding fixed operations.  We prove that
	$\bind{\mathbf{V}}{\tcurr} =
	\fixed{\sigma}_{\fixed{\theta}}(\bind{\mathbf{R}}{\tcurr})$ holds at all $\tcurr \in \timeDom$
	for $\mathbf{V} = \{x | \exists r \in \mathbf{R}(x.\mathbf{A} = r.\mathbf{A} \wedge x.RT = (r.RT \gv{\wedge} \gv{\theta}(r))
	\wedge x.RT \neq \emptyset)\}$.

	\vspace*{-7pt}
	{\footnotesize
		\begin{align*}
		& \bind{\mathbf{V}}{\tcurr}
		=
		\fixed{\sigma}_{\fixed{\theta}}(\bind{\mathbf{R}}{\tcurr})
		\\[5pt] & \Leftrightarrow
		\begin{aligned}[t]
		& \bind{\{x | \exists r \in \mathbf{R}( x.\mathbf{A} =
			r.\mathbf{A} \wedge
			x.RT = (r.RT
			\gv{\wedge} \gv{\theta}(r)) \wedge x.RT \neq
			\emptyset)\}}{\tcurr} \\ & = \{u | u \in
		\bind{\mathbf{R}}{\tcurr} \wedge \fixed{\theta}(u) \}
		\end{aligned}
		\\[5pt] & \Rightarrow^* \hspace*{-1.5pt}
		\begin{aligned}[t]
		& \{u | \exists r \in \mathbf{R}(u.\mathbf{A} = \bind{r.\mathbf{A}}{\tcurr}
		\wedge
		\tcurr \in (r.RT \gv{\wedge} \gv{\theta}(r)))
		\}
		\\ & = \{u | \exists r \in \mathbf{R}( u.\mathbf{A} = \bind{r.\mathbf{A}}{\tcurr}
		\wedge \tcurr \in r.RT
		\wedge \fixed{\theta}(\bind{r}{\tcurr}))\}
		\end{aligned}
		\\[5pt] & \Leftrightarrow
		\begin{aligned}[t]
		& \{u | \exists r \in \mathbf{R}( u.\mathbf{A} = \bind{r.\mathbf{A}}{\tcurr}
		\wedge \tcurr \in r.RT \wedge \bind{\gv{\theta}(r)}{\tcurr})\}
		\\ & = \{u | \exists r \in \mathbf{R}(
		u.\mathbf{A} = \bind{r.\mathbf{A}}{\tcurr} \wedge
		\tcurr \in r.RT \wedge \bind{\gv{\theta}(r)}{\tcurr})\}
		\end{aligned}
		\end{align*}
	}
	$^*$ The bind operator eliminates tuples with an empty reference
time and therefore ensures $RT \neq \emptyset$.
\end{IEEEproof}
}

As an example, selection $\sigma_{\gv{\theta}}(\mathbf{R})$
selects a tuple $r \in \mathbf{R}$ by restricting the tuple's
reference time $RT$. The reference time of the tuple is set to
$r.RT \wedge \theta(r)$, i.e., the intersection of the reference
time of the original tuple ($r.RT$) and the reference times when
predicate $\theta(r)$ is satisfied.
To restrict $RT$ with an ongoing boolean, we convert a tuple's
reference time into the set $S_t$ of an ongoing boolean
$\ongoingBoolVar{S_t}{S_f}$ and calculate the conjunction between the
ongoing booleans to determine the reference time of the result tuple.

\begin{example}
  Consider ongoing relation $\mathbf{X}$ with tuple
  $x = (500, \text{Spam filter}, [\dts{1}{25},\NowS),
  \rtattribute{(-\infty, \dts{8}{16})})$ and selection
  $Q = \gv{\sigma}_\theta(\mathbf{X})$ with
  $\theta = VT\ \gv{\text{overlaps}}$ $[\dts{1}{20},\dts{8}{18})$.
  Query $Q$ selects input tuple $x$ at the reference times when it
  belongs to the instantiated input relations (up to reference time
  $\dts{8}{15}$) \emph{and} when predicate $\theta(x)$ evaluates to
  \emph{true}.  The result of predicate $\theta(x)$ is ongoing boolean
  $\ongoingBool{[\dts{1}{26}, \infty)}{(-\infty, \dts{1}{26})}$. The
  reference time of result tuple $y$ is $x.RT \wedge \theta(x)$:

  \vspace*{-7pt} {\footnotesize
    \begin{align*}
      y.RT & = x.RT \wedge \theta(x) \\
           & =\rtattribute{(-\infty, \dts{8}{16})} \wedge
               \ongoingBool{[\dts{1}{26}, \infty)}{(-\infty, \dts{1}{26})}\\
           & = \ongoingBool{(-\infty, \dts{8}{16})}{[\dts{8}{16}, \infty)}
      \\   & \hspace*{0.4cm}
             \wedge \ongoingBool{[\dts{1}{26}, \infty)}{(-\infty, \dts{1}{26})} \\
           & = \ongoingBool{[\dts{1}{26}, \dts{8}{16})}{(-\infty, \dts{1}{26}), [\dts{8}{16}, \infty)} \\
           & = \rtattribute{[\dts{1}{26}, \dts{8}{16})}
    \end{align*}
    \normalsize } Thus, for selection $Q$ on input tuple $x$ we get
  result tuple:
  \[
    y = (500, \text{Spam filter}, [\dts{1}{25},\NowS),
    \rtattribute{[\dts{1}{26},\dts{8}{16})})
  \]
\end{example}

Predicates on fixed attributes retain their standard behavior.  If a
predicate on fixed attributes evaluates to \emph{true}, the result
tuple's reference time does not change as it is restricted by the
conjunction with ongoing boolean
$\ongoingBool{(-\infty, \infty)}{\emptyset}$ ($\equiv$ \emph{true}).
If a predicate evaluates to \emph{false}, the result tuple is omitted
as the conjunction with ongoing boolean
$\ongoingBool{\emptyset}{(-\infty, \infty)}$ ($\equiv$ \emph{false})
results in an empty reference time.

\section{Implementation}
\label{sec:implementation}

This section describes the implementation of ongoing data types in
the kernel of PostgreSQL. Our implementation is space-efficient and
optimized for evaluating the operations in \Cref{sec:gvOperations}.

\paragraph*{Ongoing Time Data Types} Our implementation supports
ongoing time points with the two granularities offered by PostgreSQL:
\emph{dates} with a granularity of days and \emph{timestamps} with a
granularity of microseconds.
The PostgreSQL \emph{date} and \emph{timestamp} data types are
extended to structures composed of two fixed dates and two fixed
timestamps, respectively, to represent ongoing time points
$\minPoint{a}{b}$.  Time point $\NowS$ is represented as
$\minPoint{-\infty}{\infty}$.  Note that PostgreSQL natively provides
representations for $-\infty$ and $\infty$ as fixed dates and
timestamps.  The extensions of the \emph{date} and \emph{timestamp}
data types also yield support for ongoing time intervals of
$\ongoingDom \times \ongoingDom$ as \emph{daterange}s and
\emph{tsrange}s in PostgreSQL.

\paragraph*{Reference Time RT} We represent a tuple's reference time
as a list of fixed time intervals. For the list, we use the built-in,
variable-length data type \emph{array} to leverage the built-in
storage, indexing, and fetching mechanisms for variable length
data types.  Its variable length guarantees that PostgreSQL allocates
the minimal amount of space to store the list of reference time
intervals.

\paragraph*{Ongoing Booleans} We represent an ongoing boolean
$\ongoingBoolVar{S_t}{S_f} \in \boolDom$ with the set $S_t$ of
reference times when the ongoing boolean is \emph{true}.  $S_t$ is
represented with the same data type as a tuple's reference time.  This
is beneficial when restricting a tuple's reference time: the logical
conjunction of a predicate and the tuple's reference time can then be
directly computed (cf.\ \Cref{sec:oper-ongo-relat}).  The time
intervals used for $S_t$ are maximal, non-overlapping, and sorted in
ascending order. These properties yield an efficient implementation of
the logical connectives with a sweep-line algorithm
(cf. \Cref{alg:logicalAnd}).

We developed new algorithms for $\gv{<}$, $\gv{\wedge}$, $\gv{\vee}$,
and $\gv{\neg}$.  The less-than predicate minimizes the number of
value comparisons and the implementation of the logical connectives
processes each time interval just once. The other operations are implemented
with the equivalences in \Cref{sec:gvOperations}.

\paragraph*{Less-Than Predicate}
The less-than predicate for ongoing time points is implemented
according to the case distinction in
  \Cref{th:coreOpsEquivalences}. The result of the less-than
predicate is an ongoing boolean, which we represent as an array of
time intervals for $S_t$ as described above.
Since an ongoing time point $\minPoint{a}{b}$ ensures $a \leq b$, we
use the decision tree in \Cref{fig:lessThanDecisionTree} to determine
the correct case with at most three comparisons.

\begin{figure}[!htb]
	\centering
	\begin{tikzpicture}[shorten >=1pt, auto, node distance = 3cm,
	thick,scale=1, every node/.style={scale=0.77}]
   \tikzstyle{dbOngoingLines}=[dkgreen!70, thick]

	\tikzstyle{frame2}=[rectangle, thick, draw=white!90!black, fill=white]

	\tikzstyle{frame}=[rectangle, ultra thick, draw=white!90!black, fill=white!90!black]

	\draw[->] (2.3, 1.9) -- (0.75, 1.2);
	\draw (1.6, 1.7) node[left] {\ok};
	\draw[->] (2.7, 1.9) -- (4.25, 1.2);
	\draw (3.8, 1.7) node[left] {\fail};
	\draw (2.5, 2) node[frame] {$b < d$};

	\draw[->](0.55, 0.9) -- (0, 0.2);
	\draw (0.35,0.65) node[left] {\ok};
	\draw[->](0.95, 0.9) -- (1.5, 0.2);
	\draw (1.2, 0.65) node[right] {\fail};
	\draw (0.75, 1) node[frame] {$b < c$};

	\draw[->](4.05, 0.9) -- (3.5, 0.2);
	\draw (3.85,0.65) node[left] {\ok};
	\draw[->](4.5, 0.9) -- (5, 0.2);
	\draw (4.7, 0.65) node[right] {\fail};
	\draw (4.25, 1) node[frame] {$a < c$};

	\draw (-0.2, 0) node[frame2] {$\representOngoingBoolean{(-\infty,\infty)}$};

	\draw[->](1.3, -0.1) -- (0.75, -0.8);
	\draw (1.1,-0.35) node[left] {\ok};
	\draw[->](1.7, -0.1) -- (2.25, -0.8);
	\draw (1.95, -0.35) node[right] {\fail};
	\draw (1.5, 0) node[frame] {$a < c$};
	\draw (-0.15, -1) node[frame2] {$\representOngoingBoolean{(-\infty, c), [b+1, \infty)}$};
	\draw (2.25, -1) node[frame2] {$\representOngoingBoolean{[b + 1, \infty)}$};

	\draw (3.5, 0) node[frame2] {$\representOngoingBoolean{(-\infty, c)}$};

	\draw (5, 0) node[frame2] {$\representOngoingBoolean{\ }$};

	\end{tikzpicture}
	\caption{Decision tree for $\minPoint{a}{b} \gv{<} \minPoint{c}{d}$.}
	\label{fig:lessThanDecisionTree}
\end{figure}

\paragraph*{Logical Connectives} We use a sweep-line algorithm to
implement the logical connectives. The implementation requires and
guarantees arrays with non-overlapping time intervals that are sorted
in ascending order. The implementation has the following three
properties that make it efficient:
\begin{enumerate*}
\item no sorting is required since a sweep-line algorithm guarantees
  sorted results at no cost,
\item each time interval of the input ongoing booleans is processed at
  most once, which minimizes the number of time intervals to be
  compared, and
\item the implementation minimizes the overall number of time point
  comparisons.
\end{enumerate*}
Note that the logical connectives are not only used in predicates
  but also to calculate a tuple's reference time in a relational
  algebra operator (cf. \Cref{th:raOperators}).
\Cref{alg:logicalAnd} shows the implementation of the logical
conjunction. The efficient implementation of the conjunction is
important since the conjunction is used to calculate a result
  tuple's reference time in all relational algebra operators.

\SetAlCapSkip{3pt}
\IncMargin{1em}
\begin{algorithm}[h] \footnotesize

  \SetKwInput{KwFunction}{Procedure}
  \KwFunction{Conjunction $\mathbf{b_1} \gv{\wedge} \mathbf{b_2}$}
  \KwIn{$\mathbf{b_1}, \mathbf{b_2} \in \boolDom$:
        two arrays of non-overlapping time intervals in ascending order}
  \KwOut{$\mathbf{b_r} \in \boolDom$: array of non-overlapping time intervals in ascending order}
  \vspace{5pt}

  $\mathbf{b_r} = \representOngoingBoolean{\ }$;\hspace{5pt}
  $i_1 \leftarrow \mathbf{b_1}$.first; \hspace{5pt}
  $i_2 \leftarrow \mathbf{b_2}$.first\;

  \While{$i_1 \neq \text{\emph{nil}} \wedge i_2 \neq \text{\emph{nil}}$ \label{lineAnd:emptiness}}{
    \lIf{$i_1.t_e \leq i_2.t_s$}{$i_1 \leftarrow \mathbf{b_1}$.next} \label{lineAnd:fetchB1}
    \lElseIf{$i_2.t_e \leq i_1.t_s$}{$i_2 \leftarrow \mathbf{b_2}$.next} \label{lineAnd:fetchB2}
    \Else {\tcp{append intersection of $i_1$ and $i_2$}$\mathbf{b_r}$.append$([\max(i_1.t_s, i_2.t_s), \min(i_1.t_e, i_2.t_e)))$\; \label{lineAnd:resultAppend}
            \leIf{$i_1.t_e < i_2.t_e$}
                 {$i_1 \leftarrow \mathbf{b_1}$.next}
                 {$i_2 \leftarrow \mathbf{b_2}$.next}
           } \label{lineAnd:next}
  }
  \Return $\mathbf{b_r}$\; \label{lineAlg:return}

  \caption{ Conjunction on ongoing booleans.}
  \label{alg:logicalAnd}
\end{algorithm}
\DecMargin{1em}

\paragraph*{Query Optimization}

For the relational operators on ongoing relations, the same
rules hold as for the relational algebra operators on
fixed relations.  For instance, the equivalence
$\sigma_{\theta_{1} \wedge \theta_{2}}(\mathbf{R}) \equiv
\sigma_{\theta_1} (\sigma_{\theta_2} (\mathbf{R}))$ holds for an
ongoing relation $\mathbf{R}$.  After the rewriting, existing
optimization techniques, such as selection push-down, join ordering,
and cost-based selection of evaluation algorithms, can be used.

To leverage database optimization strategies and algorithms for
queries on ongoing relations, we split a conjunctive predicate into
a conjunctive predicate over fixed attributes only and a conjunctive
predicate that references ongoing attributes.  The predicate over
fixed attributes does not depend on the reference time and can
therefore be evaluated in the where clause.  The predicate over
ongoing attributes is used in the calculation of the result
tuple's reference time (cf. \Cref{th:raOperators}).


\begin{table*}[!htb]
	\caption{Characteristics of the experiment data sets.}
	\label{tab:evalDatasets}
	\centering
	\begin{tabular}{l|cccc|ccc}
		\hline
		& \multicolumn{3}{c}{\textbf{MozillaBugs}} & \multirow{2}{*}{\textbf{Incumbent}} & \multirow{2}{*}{$\mathbf{D^{ex}}$}
		& \multirow{2}{*}{$\mathbf{D^{sh}}$} & \multirow{2}{*}{$\mathbf{D^{sc}}$} \\
		& \textbf{BugInfo} $\mathbf{B}$ & \textbf{BugAssignment} $\mathbf{A}$ & \textbf{BugSeverity} $\mathbf{S}$ & & & & \\
		\hline
		Cardinality & 394,878 & 582,668 & 434,078 & 83,852 & 10M & 10M & 35M \\
		\# ongoing & 60,372 (15\%) & 63,588 (11\%) & 61,113 (14\%) & 15,805 (19\%) & 15\% & 15\% & 20\% \\
		Time intervals & $[a, \NowS)$ & $[a, \NowS)$ & $[a, \NowS)$  & $[a, \NowS)$ & $[a, \NowS)$ & $[\NowS, b)$
		& $[a, \NowS)$ \\
		Time span & 20 years & 20 years & 20 years & 16 years & 10 years & 10 years & 10 years \\
		\hline
	\end{tabular}
\end{table*}

\section{Evaluation}
\label{sec:evaluation}

This section compares runtime, result size, and storage requirements
of our solution with the state-of-the-art
solution from Clifford et al.\ \cite{now} and Torp et
al.~\cite{modificationSemanaticsNow}.  We vary the temporal predicate
as well as the location of ongoing time intervals to
evaluate their effects on runtime and result size.

\subsection{Setup}
\label{sec:eval-setup}

The empirical evaluation is conducted on a 3.40 GHz machine with 16GB
main memory and an SSD. The client and the database server run on the
same machine. We use the PostgreSQL 9.4.0 kernel extended with our
implementation of ongoing data types and the operations on them.

\Cref{tab:evalDatasets} summarizes the real-world and synthetic data
sets. As ongoing time intervals we use expanding time intervals
$[a, \NowS)$ and shrinking time intervals $[\NowS, b)$.  Note that the
duration of expanding ongoing time intervals increases as the
reference time increases.
The earlier an expanding time interval starts, the more time intervals
it overlaps with.  We use the real-world data sets
\emph{MozillaBugs}~\cite{bugdatasets} and
\emph{Incumbent}~\cite{incumbent}.  The \emph{MozillaBugs} data set
records the history of bugs in the Mozilla project. It contains
  the following three relations.  (1) \textbf{BugInfo} records general
  information about a bug: ID, product, component, operating system,
  textual description, and valid time. Bugs that have not been
resolved as of the date of the data export have ongoing valid time
intervals. (2) \textbf{BugAssignment} records the email address
  of the person assigned to a bug, the bug id, and the valid time.
  (3) \textbf{BugSeverity} records the bug id, the severity of the
  bug, and the valid time.  The last assignment and last severity of
  bugs with ongoing valid times have ongoing valid times as well.
\emph{Incumbent} records the
valid time periods during which projects are assigned to university
employees.
We converted project assignments that were not finished at the date
of the data export into tuples with ongoing assignments, resulting
in 19\% ongoing tuples.

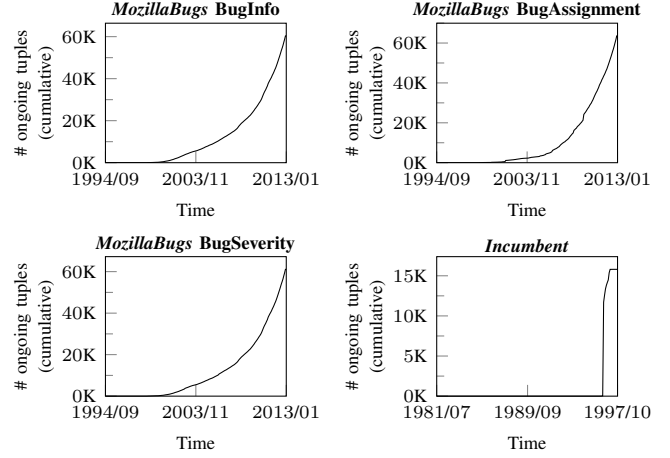
\begin{figure}[!htb] \centering
  \begin{tikzpicture}
    \tikzstyle{every node}=[font={\fontsize{7}{9}\selectfont}]
    \begin{groupplot}[
      group style = {group size = 2 by 2,horizontal sep=2cm, vertical sep=1.25cm},
      width=0.45\columnwidth,
      ylabel style={align=center},
      xlabel={Time},
      ylabel={\# ongoing tuples \\ (cumulative)},
      legend pos=outer north east,
      legend style = {cells={anchor=west}},
      minor y tick num=1,
      tick pos=left,
      scaled y ticks=base 10:-3,
      ytick scale label code/.code={},
      yticklabel={\pgfmathprintnumber{\tick}K},
      ylabel shift = -4pt,
      xmin=0,
      ymin=0,
    ]
    \nextgroupplot [
	  title style={yshift=-8pt},
	  title={\textbf{\emph{MozillaBugs}} \textbf{BugInfo}},
     xtick={0,110,220},
     xticklabels={$\dtsYear{1994}{9}$,$\dtsYear{2003}{11}$,$\dtsYear{2013}{1}$},
     xmax=220,
    ]
    \addplot[solid, color=black]
      table[x expr=\coordindex, y=count]
      {./experimentData/histograms/cumulativeMozillaComplexOngoingBugs.data};

  	\nextgroupplot [
      title style={yshift=-8pt},
      title={\textbf{\emph{MozillaBugs}} \textbf{BugAssignment}},
      xtick={0,110,220},
      xticklabels={$\dtsYear{1994}{9}$,$\dtsYear{2003}{11}$,$\dtsYear{2013}{1}$},
      xmax=220,
      ]
      \addplot[solid, color=black]
      table[x expr=\coordindex, y=count]
      {./experimentData/histograms/cumulativeMozillaComplexOngoingAssigned.data};

     \nextgroupplot [
      title style={yshift=-8pt},
      title={\textbf{\emph{MozillaBugs}} \textbf{BugSeverity}},
      xtick={0,110,220},
      xticklabels={$\dtsYear{1994}{9}$,$\dtsYear{2003}{11}$,$\dtsYear{2013}{1}$},
      xmax=220,
      ]
      \addplot[solid, color=black]
      table[x expr=\coordindex, y=count]
      {./experimentData/histograms/cumulativeMozillaComplexOngoingSeverity.data};

  \nextgroupplot [
      title style={yshift=-8pt},
      title={\textbf{\emph{Incumbent}}},
      xtick={0,98,195},
      xticklabels={$\dtsYear{1981}{7}$, $\dtsYear{1989}{9}$, $\dtsYear{1997}{10}$},
      xmax=195,
      ]
      \addplot[solid, color=black]
      table[x expr=\coordindex, y=count]
      {./experimentData/histograms/cumulativeIncumbentOngoing.data};
    \end{groupplot}
  \end{tikzpicture}
  \caption{Start point distribution of ongoing intervals.}
  \label{fig:datasetsOngoingDistribution}
\end{figure}

\Cref{fig:datasetsOngoingDistribution} shows the distribution of the
start points of the ongoing time intervals. In \emph{MozillaBugs},
50\% of the tuples with ongoing time intervals in relations \textbf{BugInfo},
  \textbf{BugAssignment}, and \textbf{BugSeverity} are located within the last two years
of the history.  In \emph{Incumbent}, all ongoing project assignments
started within the last year of the history.
For experiments with an increasing number of tuples we grow the size
of the real-world data sets by growing the history backward.
This means that the percentage of ongoing time intervals decreases as
the data size grows. For \emph{MozillaBugs}, we grow the history
  backward for the \textbf{BugInfo} relation and use all records in
  the other two relations that match to the bug ids in
  \textbf{BugInfo}.

To maximize performance we implemented the bind operator of Clifford
et al.~\cite{now} in the PostgreSQL 9.4.0 kernel as a C
function that is called when an ongoing attribute is accessed~\cite{modificationSemanaticsNow}.
$\text{Cliff}_{\max}$ refers to Clifford's approach that uses a
reference time that is greater than the latest end point.
It represents the typical use case with reference times close to the
current time.

We use two relational algebra operators for the evaluation: selection
$Q^{\sigma}_i = \sigma_{VT\ \text{pred}_i\ [t_s, t_e)}(\mathbf{R})$
with a temporal predicate on the valid time and join
$Q^\Join_i = \mathbf{R} \gv{\Join}_{\theta_{N} \wedge\ \mathbf{R}.VT \
  \text{pred}_i\ \mathbf{S}.VT} \mathbf{S}$ whose join predicate
includes equality predicates on non-temporal attributes ($\theta_{N}$)
and a temporal predicate $\text{pred}_i$ on the valid
time. $\mathbf{S}$ and $\mathbf{R}$ refer to the same relation. The
fixed time interval $[t_s, t_e)$ in the selection predicate spans the
last 10\% of the data history.  Selection is a fast operator and will
show the overhead of our approach; join queries are common for
database systems and representative for different workloads. On
  \emph{MozillaBugs}, we use a complex join query to evaluate our
  approach on a heavier workload as well. The join query determines
  for a person similar bugs that are open at any time when the person
  is working on a bug with severity \emph{major}.  Similar bugs are
  bugs that affect the same product, component, and operating system
  ($\theta_{sim}$):
	\begin{align*}
    QC^{\Join}_{i} = \mathbf{A}
    & \Join_{\mathbf{A}.ID = \mathbf{S}.ID \wedge \mathbf{A}.VT
      \text{ overlaps } \mathbf{S}.VT \wedge
      \text{Severity} = \text{'major'}} \mathbf{S}
    \\
    & \Join_{\mathbf{A}.ID = \mathbf{B}.ID} \mathbf{B}
      \Join_{\theta_{sim} \wedge  \mathbf{A}.VT \
      \text{pred}_i\ \mathbf{B'}.VT} \mathbf{B'}
	\end{align*}
 As temporal predicates, we use \emph{overlaps}
($\text{pred}_{\thetaOverlaps}$) and \emph{before}
($\text{pred}_{\thetaBefore}$). These predicates are representative
for the most commonly used temporal predicates \cite{tpch,oip, dip,
  pij, intervalJoinHelmer}.
The ongoing approach uses the predicates for ongoing time intervals (cf. \Cref{sec:gvOperations}). To maximize the performance of Clifford's approach, we use the predicates for fixed time intervals.

\subsection{Query Re-Evaluations}
\label{sec:eval:queryEvaluation}

Our approach evaluates a query to an \emph{ongoing result} that is
returned to an application.  Since ongoing results do not get
invalidated by time passing by, the application does not have to
re-evaluate the query.  In contrast, Clifford's query results get
invalidated as time passes by and thus, the application must
re-evaluate the query. First, we evaluate the break-even point of the
ongoing approach for different predicates. Next, we evaluate the
impact of the location and number of ongoing time intervals on the
runtime.

\paragraph*{Number of Query Re-Evaluations}

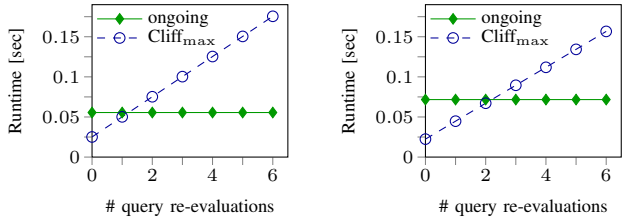
\begin{figure}[!htb]\centering
  \begin{subfigure}{0.45\columnwidth}
		\begin{tikzpicture}[font={\fontsize{7}{9}\selectfont}]
		\begin{groupplot}[
		group style = {group size = 2 by 2,
			horizontal sep=2cm},
		width=1.05\columnwidth,
		legend pos=north west,
		legend entries = {ongoing,Cliff$_{\max}$},
		legend style = {draw=none,
			font={\fontsize{7}{9}\selectfont},
			row sep=-3pt,
			inner xsep=0pt,
			inner ysep=0pt,
			fill=none,
			at={(0.02,1)},
			cells={anchor=west}},
		ymin = 0,
		xlabel={\# query re-evaluations},
		tickwidth=1mm,
		tick pos=left,
		xtick style = {yshift=-3pt},
		xticklabel style = {yshift=-1pt},
		ytick style = {xshift=-3pt},
		yticklabel style = {xshift=-1pt},
		minor tick num=1,
		scaled y ticks=base 10:-3,
		ytick scale label code/.code={},
		yticklabel style={
			/pgf/number format/fixed,
			/pgf/number format/precision=3,
		},
		]
		\nextgroupplot [
		ylabel={Runtime [sec]},
		xmin=0,
		xmax=6.5,
		xtick={0,2,...,6},
		ymax=190,
		]

		\addplot[mark=diamond*, mark options={solid}, solid, color=dkgreen]
		table[x=numRefreshes, y=matViewOngoing90.0]
		{./experimentData/ongoingResult/comparison_ongoing_reevaluation_incumbent_overlaps_19940806_19970303.data};
		\addplot[mark=o, mark options={solid}, dashed, color=darkblue]
		table[x=numRefreshes, y=matViewBind1998-01-0190.0]
		{./experimentData/ongoingResult/comparison_ongoing_reevaluation_incumbent_overlaps_19940806_19970303.data};

		\end{groupplot}
		\end{tikzpicture}
		\caption{$Q^\sigma_\thetaOverlaps$ with \emph{overlaps}.}
		\label{fig:eval_numRefreshesSelection:overlaps}
	\end{subfigure}
	\hspace*{0.2cm}
	\begin{subfigure}{0.45\columnwidth}
		\begin{tikzpicture}[font={\fontsize{7}{9}\selectfont}]
		\begin{groupplot}[
		group style = {group size = 2 by 2,
			horizontal sep=1.5cm},
		width=1.05\columnwidth,
		legend pos=north west,
		legend entries = {ongoing,Cliff$_{\max}$},
		legend style = {draw=none,
			font={\fontsize{7}{9}\selectfont},
			row sep=-3pt,
			inner xsep=0pt,
			inner ysep=0pt,
			fill=none,
			at={(0.02,1)},
			cells={anchor=west}},
		ymin = 0,
		xlabel={\# query re-evaluations},
		tickwidth=1mm,
		tick pos=left,
		xtick style = {yshift=-3pt},
		xticklabel style = {yshift=-1pt},
		ytick style = {xshift=-3pt},
		yticklabel style = {xshift=-1pt},
		minor tick num=1,
		]
		\nextgroupplot [
		ylabel={Runtime [sec]},
		xmin=0,
		xmax=6.5,
		xtick={0,2,...,6},
		ymax=190,
		scaled y ticks=base 10:-3,
		ytick scale label code/.code={},
		yticklabel style={
			/pgf/number format/fixed,
			/pgf/number format/precision=3,
		},
		]

		\addplot[mark=diamond*, mark options={solid}, solid, color=dkgreen]
		table[x=numRefreshes, y=matViewOngoing90.0]
		{./experimentData/ongoingResult/comparison_ongoing_reevaluation_incumbent_before_19970401_19971201.data};
		\addplot[mark=o, mark options={solid}, dashed, color=darkblue]
		table[x=numRefreshes, y=matViewBind1998-01-0190.0]
		{./experimentData/ongoingResult/comparison_ongoing_reevaluation_incumbent_before_19970401_19971201.data};

		\end{groupplot}
		\end{tikzpicture}
		\caption{$Q^\sigma_\thetaBefore$ with \emph{before}.}
		\label{fig:eval_numRefreshesSelection:before}
	\end{subfigure}
	\vspace*{4pt}
	\caption{Number of query re-evaluations on \emph{Incumbent}.}
	\label{fig:eval_numRefreshesSelection}
\end{figure}

The ongoing approach has a runtime overhead due to the handling of the
predicates on ongoing time points and time intervals and due to
possibly larger result sizes (cf.\ \Cref{sec:evaluation:storage}).
This is shown in \Cref{fig:eval_numRefreshesSelection} on the real
world data \emph{Incumbent} for the temporal predicates
\emph{overlaps} and \emph{before}.  Clearly, the ongoing approach
already performs better after very few query re-evaluations.
Specifically, the ongoing approach is faster after two re-evaluations
for the \emph{overlaps} predicate
(\Cref{fig:eval_numRefreshesSelection:overlaps}) and after three
re-evaluations for the \emph{before} predicate
(\Cref{fig:eval_numRefreshesSelection:before}).  Selection
$Q^\sigma_\thetaOverlaps$ is faster than selection
$Q^\sigma_\thetaBefore$ for ongoing time intervals because the
optimized implementation of the \emph{overlaps} predicate requires
about half as many fixed-value comparisons per tuple as the
\emph{before} predicate.

\paragraph*{Location of Ongoing Time Intervals}

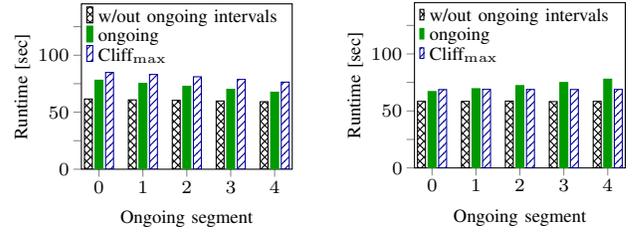
\begin{figure}[!htb]
	\centering
  \begin{subfigure}{0.45\columnwidth}
    \begin{tikzpicture}[font={\fontsize{7}{9}\selectfont}]
    \pgfplotsset{compat=1.11,
      /pgfplots/ybar legend/.style={
        /pgfplots/legend image code/.code={
          \draw[##1,/tikz/.cd,yshift=-0.25em] (0cm,0cm) rectangle (3pt,0.8em);},
      },
    }
    \begin{groupplot}[
      group style = {group name=locationPlot, group size = 2 by 1,
              horizontal sep=1.5cm},
      width=1.1\columnwidth,
      legend pos=north west,
      legend entries = {w\textrm{/}out ongoing intervals,ongoing,Cliff$_{\max}$},
      legend style = {draw=none,
              font={\fontsize{7}{9}\selectfont},
              row sep=-3pt,
              inner xsep=0pt,
              inner ysep=0pt,
              fill=none,
              at={(0.02,1)},
              cells={anchor=west}},
      legend image post style={scale=0.75},
      ylabel={Runtime [sec]},
      ylabel shift = -4pt,
      xlabel={Ongoing segment},
      tickwidth=1mm,
      tick pos=left,
      minor y tick num=1,
      xtick style = {yshift=0pt},
      xticklabel style = {yshift=2pt},
      ytick style = {xshift=-3pt},
      yticklabel style = {xshift=-1pt},
      xtick = {0,1,2,3,4},
    ]

      \nextgroupplot [
        ybar=2.5*\pgflinewidth,
        bar width=3pt,
        enlarge x limits=true,
        ymax=145,
        ymin=0,
      ]
    \addplot[solid, color=black, pattern=crosshatch]
      table[x=size, y=base2]
      {./experimentData/ongoingResult/comparison_ongoing_result_runtime_syntheticLocation_expanding_overlaps_join.data};
      \addplot[solid, color=dkgreen, fill=dkgreen]
        table[x=size, y=ongoing]
          {./experimentData/ongoingResult/comparison_ongoing_result_runtime_syntheticLocation_expanding_overlaps_join.data};
      \addplot[solid, color=darkblue, pattern=north east lines, pattern color=darkblue]
        table[x=size, y=bind42020-01-01]
          {./experimentData/ongoingResult/comparison_ongoing_result_runtime_syntheticLocation_expanding_overlaps_join.data};
    \end{groupplot}
  \end{tikzpicture}
  \caption{$Q^\Join_\thetaOverlaps$ on $D^{ex}$.}
  \label{fig:eval_location_expanding}
	\end{subfigure}
\hspace*{0.2cm}
	\begin{subfigure}{0.45\columnwidth}
		\begin{tikzpicture}[font={\fontsize{7}{9}\selectfont}]
		\pgfplotsset{compat=1.11,
			/pgfplots/ybar legend/.style={
				/pgfplots/legend image code/.code={
					\draw[##1,/tikz/.cd,yshift=-0.25em] (0cm,0cm) rectangle (3pt,0.8em);},
			},
		}
		\begin{groupplot}[
		group style = {group name=locationPlot, group size = 2 by 1,
			horizontal sep=1.5cm},
		width=1.1\columnwidth,
		legend pos=north west,
		legend entries = {w\textrm{/}out ongoing intervals,ongoing,Cliff$_{\max}$},
		legend style = {draw=none,
			font={\fontsize{7}{9}\selectfont},
			row sep=-3pt,
			inner xsep=0pt,
			inner ysep=0pt,
			fill=none,
			at={(0.02,1)},
			cells={anchor=west}},
		legend image post style={scale=0.5},
		ylabel={Runtime [sec]},
		ylabel shift = -4pt,
		xlabel={Ongoing segment},
		tickwidth=1mm,
		tick pos=left,
		minor y tick num=1,
		xtick style = {yshift=0pt},
		xticklabel style = {yshift=2pt},
		ytick style = {xshift=-3pt},
		yticklabel style = {xshift=-1pt},
		xtick = {0,1,2,3,4},
		]
		\nextgroupplot [
		ybar=2.5*\pgflinewidth,
		bar width=3pt,
		enlarge x limits=true,
		ymin=0,
		ymax=145,
		]
        \addplot[solid, color=black, pattern=crosshatch]
		table[x=size, y=base2]
		{./experimentData/ongoingResult/comparison_ongoing_result_runtime_syntheticLocation_shrinking_overlaps_join.data};
		\addplot[solid, color=dkgreen, fill=dkgreen]
		table[x=size, y=ongoing]
		{./experimentData/ongoingResult/comparison_ongoing_result_runtime_syntheticLocation_shrinking_overlaps_join.data};
		\addplot[solid, color=darkblue, pattern=north east lines, pattern color=darkblue]
		table[x=size, y=bind42020-01-01]
		{./experimentData/ongoingResult/comparison_ongoing_result_runtime_syntheticLocation_shrinking_overlaps_join.data};
		\end{groupplot}
		\end{tikzpicture}
		\caption{$Q^\Join_\thetaOverlaps$ on $D^{sh}$.}
		\label{fig:eval_location_shrinking}
	\end{subfigure}
	\vspace*{4pt}
	\caption{Location of ongoing time intervals.}
	\label{fig:eval_location}
\end{figure}

We vary the location of the
ongoing time intervals by dividing the 10 year history into 5 segments
(2 years each) and placing all start points ($D^{ex}$) or end points
($D^{sh}$) of the ongoing intervals into one of the segments. Ongoing
segment 0 spans the first two years. \Cref{fig:eval_location} shows the impact of the
location on the runtime for one re-evaluation.
Since $D^{ex}$ contains expanding ongoing time intervals, the runtime
of the ongoing approach decreases for the \emph{overlaps} predicate if
the ongoing time intervals are placed in the later segments (cf.\ \Cref{fig:eval_location_expanding}).
\Cref{fig:eval_location_shrinking} shows that the opposite observation holds
for shrinking ongoing time intervals in $D^{sh}$ since their duration is longer when their end points are placed in later ongoing segments.
To establish a baseline for the runtime, we replaced all ongoing time
intervals in the two datasets with fixed time intervals and evaluated
query $Q^\Join_\thetaOverlaps$ on these data sets (without ongoing
time intervals).
Observe that the baseline runtime accounts for 80\% to 90\% of the runtime of the
ongoing approach.  Thus, the join processing is the expensive part and
the runtime overhead for processing ongoing time intervals is less
than 20\%.

\paragraph*{Number of Input Tuples}

We evaluate the scalability by increasing the size of the input
relation. \Cref{fig:eval_runtimeScalabilityRuntime} shows that the
ongoing approach has a similar linear runtime
increase as Clifford's approach does with increasing input sizes.  Thus, as shown in
\Cref{fig:eval_runtimeScalabilityRefreshes}, the number of query
re-evaluations after which the ongoing approach performs better stays
constant as the number of input tuples increases.

\begin{figure}[!htb]
	\centering
	\begin{subfigure}[b]{0.45\columnwidth}
		\begin{tikzpicture}[font={\fontsize{7}{9}\selectfont}]
		\begin{groupplot}[
		group style = {group size = 2 by 1,
			horizontal sep=1.5cm},
		width=1.05\columnwidth,
		legend pos=north west,
		legend style = {draw=none,
			font={\fontsize{7}{9}\selectfont},
			row sep=-3pt,
			inner xsep=0pt,
			inner ysep=0pt,
			fill=none,
			at={(0.02,1)},
			cells={anchor=west}},
		ymin = 0,
		ylabel={Runtime [sec]},
		ylabel shift = -3pt,
		xlabel={\# input tuples [M]},
		tickwidth=1mm,
		tick pos=left,
		xtick style = {yshift=-3pt},
		xticklabel style = {yshift=-1pt},
		ytick style = {xshift=-3pt},
		yticklabel style = {xshift=-1pt},
		minor tick num=1,
		]
		\nextgroupplot [
		legend entries = {ongoing,Cliff$_{\max}$},
		ymax=22.5,
		xmin=0,
		]
		\addplot[mark=diamond*, mark options={solid}, solid, color=dkgreen]
		table[x=size, y=ongoing]
		{./experimentData/ongoingResult/comparison_ongoing_result_runtime_syntheticScale_overlaps_20191230_20201229.data};
		\addplot[mark=o, mark options={solid}, dashed, color=darkblue]
		table[x=size, y=bind42020-01-01]
		{./experimentData/ongoingResult/comparison_ongoing_result_runtime_syntheticScale_overlaps_20191230_20201229.data};
		\end{groupplot}
		\end{tikzpicture}
		\caption{Runtime.}
		\label{fig:eval_runtimeScalabilityRuntime}
	\end{subfigure}
\hspace*{0.2cm}
	\begin{subfigure}[b]{0.45\columnwidth}
		\begin{tikzpicture}[font={\fontsize{7}{9}\selectfont}]
		\begin{groupplot}[
		width=1.05\columnwidth,
		legend pos=north west,
		legend style = { draw=none,
			font={\fontsize{7}{9}\selectfont},
			row sep=-3pt,
			inner xsep=0pt,
			inner ysep=0pt,
			fill=white,
			cells={anchor=west}},
		ymin = 0,
		xlabel={\# input tuples [M]},
		tickwidth=1mm,
		tick pos=left,
		xtick style = {yshift=-3pt},
		xticklabel style = {yshift=-1pt},
		ytick style = {xshift=-3pt},
		yticklabel style = {xshift=-1pt},
		minor tick num=1,
		]
		\nextgroupplot [
		title style={yshift=-8pt},
		ylabel={\# re-evaluations},
		ymin=0,
		ymax=3,
		xmin=0,
		]
		\addplot[mark=o, mark options={solid}, dashed, color=darkblue]
		table[x=size, y=ongoingRatio2020-01-01]
		{./experimentData/ongoingResult/comparison_ongoing_result_runtime_syntheticScale_overlaps_20191230_20201229.data};
		\end{groupplot}
		\end{tikzpicture}
	\caption{Query re-evaluations.}
	\label{fig:eval_runtimeScalabilityRefreshes}
	\end{subfigure}
	\vspace*{4pt}
	\caption{Number of input tuples ($Q^\sigma_\thetaOverlaps$ on $D^{sc}$).}
	\label{fig:eval_runtimeScalability}
\end{figure}

\subsection{Instantiated Query Results via Materialized Views}

Ongoing relations can easily be combined with materialized views to
efficiently compute instantiated results at different reference times.
This allows applications that do not want to handle ongoing relations
explicitly to leverage the performance benefits of ongoing relations.
We evaluate the runtime amortization of the ongoing approach, i.e.,
at how many different reference times $n$ an instantiated result must be returned
to an application, such that calculating the ongoing result and instantiating
it at the $n$ reference times outperforms Clifford's approach, which must
calculate the query at each of the $n$ reference times.
The main factors for the amortization are
\begin{enumerate*}[label=(\arabic*)]
\item the complexity of the query and
\item the reference time used for the instantiation.
\end{enumerate*}

\begin{figure}[!htb]
	\centering
	\begin{subfigure}{0.45\columnwidth}
		\begin{tikzpicture}[font={\fontsize{7}{9}\selectfont}]
		\begin{groupplot}[
		group style = {group size = 2 by 1,
			horizontal sep=1.5cm},
		width=1.05\columnwidth,
		ylabel style={align=center},
		xlabel={\# input bugs [k]},
		tickwidth=1mm,
		tick pos=left,
		xtick style = {yshift=-3pt},
		xticklabel style = {yshift=-1pt},
		minor tick num=1,
		ytick style = {xshift=-3pt},
		yticklabel style = {xshift=-1pt},
		xmin=50,
		]

		\nextgroupplot [
		title style={yshift=-8pt},
		ylabel={\# instantiations \\ for amortization},
		ylabel shift = -2pt,
		ymin = 1,
		ymax = 2,
		]
		\addplot[color=darkblue, mark=o, mark options={solid}, dashed] table[x=size, y=amor2014-01-01]
		{./experimentData/amortization/withBindMatView/comparison_amortization_runtime_mozilla_overlaps_selection_20110225_20121225_bindMatView.data};
		\end{groupplot}
		\end{tikzpicture}
		\caption{Selection $Q^\sigma_\thetaOverlaps(\mathbf{B})$.}
	\end{subfigure}
\hspace*{0.2cm}
	\begin{subfigure}{0.45\columnwidth}
		\begin{tikzpicture}[font={\fontsize{7}{9}\selectfont}]
		\begin{groupplot}[
		group style = {group size = 2 by 1,
			horizontal sep=1.5cm},
		width=1.05\columnwidth,
		ylabel style={align=center},
		xlabel={\# input bugs [k]},
		tickwidth=1mm,
		tick pos=left,
		xtick style = {yshift=-3pt},
		xticklabel style = {yshift=-1pt},
		minor tick num=1,
		ytick style = {xshift=-3pt},
		yticklabel style = {xshift=-1pt},
		xmin=50,
		]

		\nextgroupplot [
		title style={yshift=-8pt},
		ylabel={\# instantiations \\ for amortization},
		ylabel shift = -2pt,
		ymin = 1,
		ymax = 2,
		]

		\addplot[color=darkblue, mark=o, mark options={solid}, dotted] table[x=size, y=amor2014-01-01]
		{./experimentData/amortization/comparison_amortization_runtime_mozilla_overlapsComplex_join.data};

		\end{groupplot}
		\end{tikzpicture}
		\caption{Join $QC^\Join_\thetaOverlaps(\mathbf{A}, \mathbf{S}, \mathbf{B})$.}
		\label{fig:amortizationOperators_join}
	\end{subfigure}
	\vspace*{4pt}
	\caption{Amortization for selection and join on \emph{MozillaBugs}.}
	\label{fig:amortizationOperators}
\end{figure}
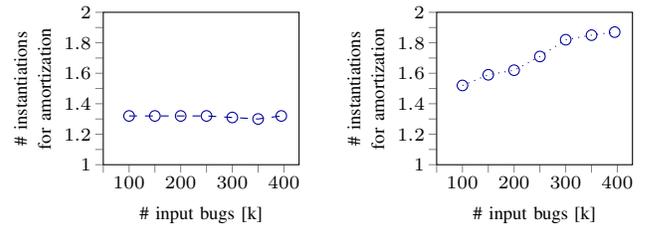

\paragraph*{Query Complexity}
\Cref{fig:amortizationOperators} shows the amortization for
selection and complex join. The number of input bugs (x-axis) is
equal to the number of tuples in relation $\mathbf{B}$
(cf. \Cref{sec:eval-setup} on how we vary the size of the
dataset). Both queries require less than two instantiations for the
amortization at all input sizes. For the selection query, the number
of reference times for amortization remains constant with varying
input size. For the complex join, it increases slightly: around 25\%
for a 300\% input bugs increase.  This is because the query
optimizer chooses a linear-time hash join for Clifford's approach
when evaluating the join with $\mathbf{B'}$, whereas it uses a
log-linear-time merge join for the ongoing approach. This additional
logarithmic component is consistent with the curve in
\Cref{fig:amortizationOperators_join}.

\paragraph*{Reference Time}

Smaller size differences of the ongoing and instantiated query result
lead to a faster runtime amortization of the ongoing approach. The
size of the ongoing result is independent of the reference time
whereas the size of the instantiated result depends on it.
\Cref{fig:eval_amortization_armotization} shows that the amortization
of the ongoing approach decreases from three
instantiations for early reference times ($\tcurr=\min$, i.e.,
smallest time point in the data set) to two instantiations for later
reference times. For the \emph{overlaps} predicate, later reference
times result in smaller size differences: the later the reference
time, the more ongoing time intervals instantiate to non-empty time
intervals.  Thus, more and more ongoing time intervals satisfy the
predicate (especially as a late selection time interval is used) and
belong to the result (\Cref{fig:eval_amortization_resultsize}).
\vspace*{-10pt}

\begin{figure}[!htb]
	\centering
		\begin{subfigure}[b]{0.45\columnwidth}
		\begin{tikzpicture}[font={\fontsize{7}{9}\selectfont}]
		\begin{groupplot}[
		group style = {group size = 2 by 1,
			horizontal sep=1.5cm},
		width=1.05\columnwidth,
		ylabel style={align=center},
		xlabel={\# input tuples [k]},
		tickwidth=1mm,
		tick pos=left,
		xtick style = {yshift=-3pt},
		xticklabel style = {yshift=-1pt},
		minor tick num=1,
		ytick style = {xshift=-3pt},
		yticklabel style = {xshift=-1pt},
		xmin=50,
		]

		\nextgroupplot [
		title style={yshift=-8pt},
		ylabel={\# instantiations \\ for amortization},
		ylabel shift = -2pt,
		ymin = 1,
		ymax = 2.75,
		legend pos=north east,
		legend entries = {$\tcurr = \dtsYear{1994}{1}$ ($\min$), $\tcurr = \dtsYear{2012}{1}$,$\tcurr = \dtsYear{2012}{9}$, $\tcurr = \dtsYear{2014}{1}$ ($\max$)},
		legend style = {draw=none,
			font={\fontsize{7}{9}\selectfont},
			row sep=-3pt,
			inner xsep=0pt,
			inner ysep=0pt,
			fill=white,
			at={(1,1.6)},
			cells={anchor=west}},
		]
		\addplot[color=darkred, mark=triangle, mark options={solid}, dotted] table[x=size, y=amor1994-01-01]
		{./experimentData/amortization/withBindMatView/comparison_amortization_runtime_mozilla_overlaps_selection_20110225_20121225_bindMatView.data};

		\addplot[color=orangeDark, mark=square, mark options={solid}, solid] table[x=size, y=amor2012-01-01]
		{./experimentData/amortization/withBindMatView/comparison_amortization_runtime_mozilla_overlaps_selection_20110225_20121225_bindMatView.data};

		\addplot[color=mauve, mark=star, mark options={solid}, dotted] table[x=size, y=amor2012-09-01]
		{./experimentData/amortization/withBindMatView/comparison_amortization_runtime_mozilla_overlaps_selection_20110225_20121225_bindMatView.data};

		\addplot[color=darkblue, mark=o, mark options={solid}, dashed] table[x=size, y=amor2014-01-01]
		{./experimentData/amortization/withBindMatView/comparison_amortization_runtime_mozilla_overlaps_selection_20110225_20121225_bindMatView.data};
		\end{groupplot}
		\end{tikzpicture}
		\caption{Amortization.}
		\label{fig:eval_amortization_armotization}
		\end{subfigure}
		\hspace*{0.2cm}
		\begin{subfigure}[b]{0.45\columnwidth}
			\begin{tikzpicture}[font={\fontsize{7}{9}\selectfont}]
			\pgfplotsset{compat=1.11,
				/pgfplots/ybar legend/.style={
					/pgfplots/legend image code/.code={
						\draw[##1,/tikz/.cd,yshift=-0.25em] (0cm,0cm) rectangle (3pt,0.8em);},
				},
			}
			\begin{groupplot}[
			group style = {group name=locationPlot, group size = 2 by 1,
				horizontal sep=1.5cm},
			width=1.05\columnwidth,
			legend pos=north west,
			legend style = {draw=none,
				font={\fontsize{7}{9}\selectfont},
				row sep=-3pt,
				inner xsep=0pt,
				inner ysep=0pt,
				fill=none,
				at={(-0.1,1.7)},
				cells={anchor=west}},
			ylabel={\# result tuples [k]},
			ylabel shift = -4pt,
			tickwidth=1mm,
			tick pos=left,
			xtick style = {yshift=-3pt},
			xticklabel style = {yshift=-1pt},
			ytick style = {xshift=-3pt},
			yticklabel style = {xshift=-1pt},
			minor tick num=1,
			ymin = 0,
			yticklabel style={
				/pgf/number format/fixed,
				/pgf/number format/precision=2
			},
			]
			\nextgroupplot [
			xlabel={\# input tuples [k]},
			xmin=50,
			]
			\addplot[color=dkgreen, mark=diamond*, mark options={solid}, solid] table[x=size, y=matView]
			{./experimentData/amortization/withBindMatView/comparison_resultsize_mozilla_overlaps_20110225_20121225_differentBindDates.data};
			\addlegendentry{ongoing}

			\addplot[color=darkblue, mark=o, mark options={solid}, dashed] table[x=size, y=bind2014-01-01]
			{./experimentData/amortization/withBindMatView/comparison_resultsize_mozilla_overlaps_20110225_20121225_differentBindDates.data};
			\addlegendentry{$\tcurr = \dtsYear{2014}{1}$ ($\max$)}

			\addplot[color=mauve, mark=star, mark options={solid}, dotted] table[x=size, y=bind2012-09-01]
			{./experimentData/amortization/withBindMatView/comparison_resultsize_mozilla_overlaps_20110225_20121225_differentBindDates.data};
			\addlegendentry{$\tcurr = \dtsYear{2012}{9}$}

			\addplot[color=orangeDark, mark=square, mark options={solid}, solid] table[x=size, y=bind2012-01-01]
			{./experimentData/amortization/withBindMatView/comparison_resultsize_mozilla_overlaps_20110225_20121225_differentBindDates.data};
			\addlegendentry{$\tcurr = \dtsYear{2012}{1}$}

			\addplot[color=darkred, mark=triangle, mark options={solid}, dotted] table[x=size, y=bind1994-01-01]
			{./experimentData/amortization/withBindMatView/comparison_resultsize_mozilla_overlaps_20110225_20121225_differentBindDates.data};
			\addlegendentry{$\tcurr = \dtsYear{1994}{1}$ ($\min$)}

			\end{groupplot}
			\end{tikzpicture}
			\caption{Result size.}
			\label{fig:eval_amortization_resultsize}
		\end{subfigure}
	\vspace*{4pt}
		\caption{Amortization for $Q^\sigma_\thetaOverlaps(\mathbf{B})$ on \emph{MozillaBugs}.}
	\label{fig:eval_amortization}
\end{figure}

\subsection{Storage}
\label{sec:evaluation:storage}

The ongoing approach requires additional storage for each
tuple and for the tuples that belong to the ongoing result but not
to Clifford's result.  The per-tuple storage overhead is the
additional $RT$ attribute and a doubling of the size of the valid
time attribute (because ongoing rather than fixed values are used).
Typically, the value of the $RT$ attribute can be represented with
one fixed time interval.

\crc{
\paragraph*{Per-Tuple Storage}
The per-tuple storage overhead consists of the additional $RT$
  attribute and doubling the size of the valid time (+8 Bytes). We
  first analyze the cardinality of the $RT$ attribute, i.e., the
  number of fixed intervals that is needed to represent a tuple's
  reference time, and then discuss the additional storage
  requirements.

	\Cref{tab:rtSizePredicates} shows that
	the result of the common predicates on ongoing time intervals (cf. \Cref{fig:remainingOps}) can be
	represented with one interval in most cases.

	\begin{table}[!htb]  \centering
		\caption{Predicates: maximum cardinality of RT.}
		\label{tab:rtSizePredicates}
		\begin{tabular}{l|ccc} \hline
			& \multicolumn{3}{c}{Ongoing time intervals} \\
			& expanding & shrinking & expanding + shrinking \\ \hline
			before & 1 & 1 & 1 \\
			starts & 1 & 1 & 1 \\
			during & 1 & 1 & 1 \\
			meets & 1 & 1 & 1 \\
			finishes & 1 & 1 & 1 \\
			equals & 1 & 1 & 1 \\
			overlaps & 1 & 1 & 2 \\ \hline
		\end{tabular}
	\end{table}

	Thus, the typical input cardinality for subsequent logical
  connectives is one.  For conjunction
  $\mathbf{b}_1 \wedge \mathbf{b}_2$ and disjunction
  $\mathbf{b}_1 \vee \mathbf{b}_2$ the worst case output cardinality
  is $|\mathbf{b}_1| + |\mathbf{b}_2|$. Negation has an output
  cardinality of
  $|\mathbf{b}_1| - 1 \leq |\neg \mathbf{b}_1| \leq |\mathbf{b}_1| +
  1$.  Conjunction is the most widely used connective in predicates
  and is used to restrict a tuple's reference time. Its typical output
  cardinality is one. Thus, the typical cardinality of $RT$ is one as
  well.

	\Cref{tab:perTupleStorage} shows the per-tuple storage requirements
  for the three base relations of the \emph{MozillaBugs} data set and
  two query results. The $RT$ attribute contributes 29 Bytes to the
  storage size of a tuple in all five relations. This corresponds to
  the typical case where a tuple's reference time is represented with
  one fixed time interval. The constant overhead for the $RT$
  attribute can be significant for small tuple sizes (+75\% for 100B)
  and gets insignificant for larger tuples (+4\% for $\geq$
  1kB). Small tuple sizes often occur in foreign key relations. Larger
  tuple sizes occur in real-world data with descriptive attributes
  (e.g., the textual description of a bug).

	\begin{table}[!htb]
		\caption{Per-tuple storage on \emph{MozillaBugs}.}
		\label{tab:perTupleStorage}
		\centering
		\fontsize{7.5}{10}\selectfont
		\setlength{\tabcolsep}{2pt}
		\begin{tabular}{l|ccc|cc}
			\hline
			& $\mathbf{B}$ & $\mathbf{A}$  & $\mathbf{S}$ & $Q^{\sigma}_{\thetaOverlaps}(\mathbf{B})$ & $QC^{\Join}_{\thetaOverlaps}$ \\
			\hline
			avg tuple size & 968B & 90B  & 86B & 968B & 2.34kB \\
			$RT$ size & 29B (3\%) & 29B (32\%) & 29B (34\%) & 29B (3\%) & 29B (1\%) \\
			$\frac{\text{ongoing}}{\text{fixed}}$ tuple size & 104\% & 167\% & 175\% & 104\% & 103\% \\
			\hline
		\end{tabular}
	\end{table}

}

The number of additional tuples that are part of the ongoing result
but not of Clifford's result depends on the reference time. Since
ongoing results combine the results at all reference times, they must
contain at least the tuples of the largest instantiated result. If the
size of the ongoing result and the largest instantiated result are
equal, the size of the ongoing result is optimal.

\begin{figure}[!htb]
	\centering
	\begin{subfigure}[t]{0.45\columnwidth}
			\centering
		 \begin{tikzpicture}
		\tikzstyle{every node}=[font={\fontsize{7}{9}\selectfont}]
		\begin{groupplot}[
		group style = {group size = 1 by 1},
		width=1.05\columnwidth,
		ylabel style={align=center},
		xlabel={Reference time},
		ylabel={\# result tuples [k]},
		legend style = {draw=none,
			font={\fontsize{7}{9}\selectfont},
			row sep=-3pt,
			inner xsep=0pt,
			inner ysep=0pt,
			fill=none,
			at={(1,1)},
			cells={anchor=west}},
		minor y tick num=1,
		tick pos=left,
		ylabel shift = -4pt,
		xmin=0,
		ymin=0,
		ymax=200,
		every axis plot/.append style={thick}
		]
		\nextgroupplot [
		title style={yshift=-8pt},
		xtick={0,117,233},
		xticklabels={$\dts{1994}{9}$,$\dts{2004}{4}$,$\dts{2014}{1}$},
		xmax=234,
		]
		\addplot[solid, color=dkgreen] table[x expr=\coordindex, y=ongoingCount]
		{./experimentData/storage/resultsize_reftime/resultsize_reftime_mozillaSelectionOverlaps.data};
		\addlegendentry{ongoing result};

		\addplot[dashed, color=black] table[x expr=\coordindex, y=bindCount]
		{./experimentData/storage/resultsize_reftime/resultsize_reftime_mozillaSelectionOverlaps.data};
		\addlegendentry{instantiated result}
		\end{groupplot}
		\end{tikzpicture}

		\caption{Selection $Q^\sigma_\thetaOverlaps(\mathbf{B})$.}
		\label{fig:eval_storage_resultsize_selectionOverlaps}
	\end{subfigure}
	\hspace*{0.2cm}
	\begin{subfigure}[t]{0.45\columnwidth}
		\centering
		 \begin{tikzpicture}
		\tikzstyle{every node}=[font={\fontsize{7}{9}\selectfont}]
		\begin{groupplot}[
		group style = {group size = 1 by 1},
		width=1.05\columnwidth,
		ylabel style={align=center},
		xlabel={Reference time},
		ylabel={\# result tuples [k]},
		legend style = {draw=none,
			font={\fontsize{7}{9}\selectfont},
			row sep=-3pt,
			inner xsep=0pt,
			inner ysep=0pt,
			fill=none,
			at={(1,1)},
			cells={anchor=west}},
		minor y tick num=1,
		tick pos=left,
		ylabel shift = -4pt,
		xmin=0,
		ymin=200,
		ymax=400,
		every axis plot/.append style={thick}
		]
		\nextgroupplot [
		title style={yshift=-8pt},
		xtick={0,117,233},
		xticklabels={$\dts{1994}{9}$,$\dts{2004}{4}$,$\dts{2014}{1}$},
		xmax=234,
		]
		\addplot[solid, color=dkgreen] table[x expr=\coordindex, y=ongoingCount]
		{./experimentData/storage/resultsize_reftime/resultsize_reftime_mozillaSelectionBefore.data};
		\addlegendentry{ongoing result};

		\addplot[dashed, color=black] table[x expr=\coordindex, y=bindCount]
		{./experimentData/storage/resultsize_reftime/resultsize_reftime_mozillaSelectionBefore.data};
		\addlegendentry{instantiated result}
		\end{groupplot}
		\end{tikzpicture}
	\caption{Selection $Q^\sigma_\thetaBefore(\mathbf{B})$.}
	\label{fig:eval_storage_resultsize_selectionBefore}
	\end{subfigure}
	\vspace*{9pt}

	\begin{subfigure}[b]{0.45\columnwidth}
	\centering
	 \begin{tikzpicture}
		\tikzstyle{every node}=[font={\fontsize{7}{9}\selectfont}]
		\begin{groupplot}[
		group style = {group size = 1 by 1},
		width=1.05\columnwidth,
		ylabel style={align=center},
		xlabel={Reference time},
		ylabel={\# result tuples [M]},
		legend style = {draw=none,
			font={\fontsize{7}{9}\selectfont},
			row sep=-3pt,
			inner xsep=0pt,
			inner ysep=0pt,
			fill=none,
			at={(1,1)},
			cells={anchor=west}},
		minor y tick num=1,
		tick pos=left,
		ylabel shift = -4pt,
		xmin=0,
		ymin=15,
		ymax=50,
		every axis plot/.append style={thick}
		]
		\nextgroupplot [
		title style={yshift=-8pt},
		xtick={0,117,233},
		xticklabels={$\dts{1994}{9}$,$\dts{2004}{4}$,$\dts{2014}{1}$},
		xmax=234,
		]
		\addplot[solid, color=dkgreen] table[x expr=\coordindex, y=ongoingCount]
		{./experimentData/storage/resultsize_reftime/resultsize_reftime_mozillaComplexJoinOverlaps.data};
		\addlegendentry{ongoing result};

		\addplot[dashed, color=black] table[x expr=\coordindex, y=bindCount]
		{./experimentData/storage/resultsize_reftime/resultsize_reftime_mozillaComplexJoinOverlaps.data};
		\addlegendentry{instantiated result}
		\end{groupplot}
		\end{tikzpicture}
	\caption{Join $QC^\Join_\thetaOverlaps(\mathbf{A}, \mathbf{S}, \mathbf{B})$.}
	\label{fig:eval_storage_resultsize_joinOverlaps}
	\end{subfigure}
	\hspace*{0.2cm}
	\begin{subfigure}[b]{0.45\columnwidth}
		\centering
			 \begin{tikzpicture}
		\tikzstyle{every node}=[font={\fontsize{7}{9}\selectfont}]
		\begin{groupplot}[
		group style = {group size = 1 by 1},
		width=1.05\columnwidth,
		ylabel style={align=center},
		xlabel={Reference time},
		ylabel={\# result tuples [M]},
		legend style = {draw=none,
			font={\fontsize{7}{9}\selectfont},
			row sep=-3pt,
			inner xsep=0pt,
			inner ysep=0pt,
			fill=none,
			at={(1,1)},
			cells={anchor=west}},
		minor y tick num=1,
		tick pos=left,
		ylabel shift = -4pt,
		xmin=0,
		ymin=55,
		ymax=80,
		every axis plot/.append style={thick}
		]
		\nextgroupplot [
		title style={yshift=-8pt},
		xtick={0,117,233},
		xticklabels={$\dts{1994}{9}$,$\dts{2004}{4}$,$\dts{2014}{1}$},
		xmax=234,
		]
		\addplot[solid, color=dkgreen] table[x expr=\coordindex, y=ongoingCount]
		{./experimentData/storage/resultsize_reftime/resultsize_reftime_mozillaComplexJoinBefore.data};
		\addlegendentry{ongoing result};

		\addplot[dashed, color=black] table[x expr=\coordindex, y=bindCount]
		{./experimentData/storage/resultsize_reftime/resultsize_reftime_mozillaComplexJoinBefore.data};
		\addlegendentry{instantiated result}
		\end{groupplot}
		\end{tikzpicture}
	\caption{Join $QC^\Join_\thetaBefore(\mathbf{A}, \mathbf{S}, \mathbf{B})$.}
	\label{fig:eval_storage_resultsize_joinBefore}
	\end{subfigure}
	\vspace*{4pt}
	\caption{Result size vs. reference time on \emph{MozillaBugs}.}
	\label{fig:eval_storage_resultsize}
\end{figure}
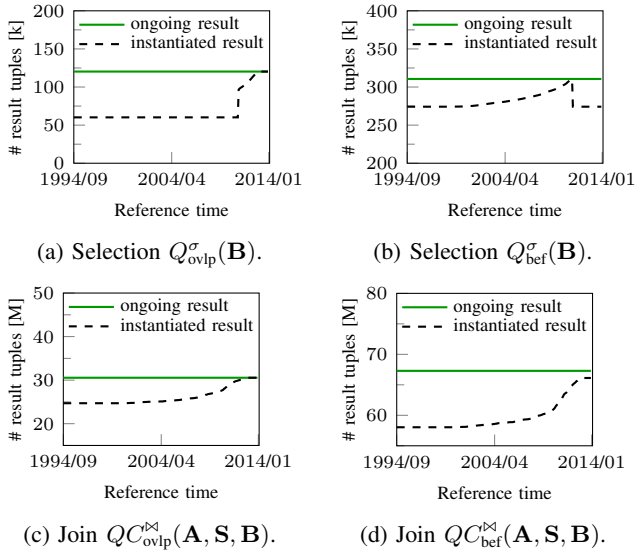

For expanding ongoing intervals the size of the ongoing result is
optimal for predicate \emph{overlaps}
(\Cref{fig:eval_storage_resultsize_selectionOverlaps} and
\Cref{fig:eval_storage_resultsize_joinOverlaps}). As the duration of
expanding time intervals increases, once an expanding time interval
overlaps with a time interval, they remain overlapping for all
reference times afterwards. Tuples are only added to the instantiated
query results with increasing reference times and thus, the ongoing
result contains exactly the tuples of the largest instantiated result.

For expanding ongoing intervals and the \emph{before} predicate, the
ongoing result reaches the optimal size for selections
(\Cref{fig:eval_storage_resultsize_selectionBefore}) and gets close to
it for joins (\Cref{fig:eval_storage_resultsize_joinBefore}).  Due to
the duration increase, expanding ongoing time intervals are before a
time interval up to a reference time and then stop being before it.
As there is one selection interval in the selection, this reference
time is the same for all expanding time intervals (it is the start
point of the selection interval). In a join, an expanding time
interval is compared to multiple time intervals. Usually there does
not exist a single reference time that belongs to the $RT$ attribute
of all result tuples, and thus, the maximum instantiated result is
smaller than the ongoing result.

\subsection{Summary}

As expected, the ongoing approach has a runtime overhead to compute
ongoing results that do not get invalidated by time passing by.  This
overhead is quite small and pays off for as little as three query
re-evaluations of Clifford's approach when returning an ongoing result
and for returning as little as two instantiated results when
leveraging the ongoing result to calculate them.
For late reference times, which are close to the current time, the
result size of the ongoing approach is equal to the result size of
Clifford's approach for the widely-used \emph{overlaps} predicate and
close to equal for other predicates.  Thus, the number of tuples that
are contained in an ongoing result but not in Clifford's result is
small.

\section{Conclusions}
\label{sec:conclusion}

We propose the first approach that evaluates queries on
ongoing relations without instantiating ongoing time points.  Ongoing
time points are preserved in query results and the results remain
valid as time passes by. For database systems this is a crucial
property as it guarantees that cached results, materialized views,
etc. have to be maintained only after explicit database modifications.
We define predicates and functions on ongoing time points and time intervals.
  We propose \emph{ongoing
  relations} that associate each tuple with a reference time
attribute. The value of the reference time attribute contains the reference times
when a tuple belongs to the instantiated relations and is restricted
by predicates on ongoing attributes.

There are several interesting topics for future research.  First, we
want to extend the set of functions for ongoing data
types to include a duration function for ongoing time intervals
whose result are ongoing integers.  Second, we plan to propose an
aggregation operator for ongoing relations and determine the
additional ongoing data types that are required to support aggregation
and group tuples in the presence of $RT$ and ongoing attributes.
Finally, we want to develop index access methods for ongoing time
points (based on the approaches for indexing fixed time intervals) and
discuss query classes that benefit from these indexes.

\balance
\bibliographystyle{IEEEtran}
\bibliography{IEEEabrv,references.bib}

\end{document}